%% file: paper.tex
\documentclass[number,sort&compress,twocolumn,5p]{elsarticle}
\usepackage[binary-units=true,separate-uncertainty=true]{siunitx}
\usepackage{lineno,hyperref}
\usepackage{tikz}
\usepackage{amssymb,amsmath,mathtools}
\usepackage{caption}
\usepackage{xspace}
\usepackage{booktabs}
\usepackage{siunitx}
\usepackage{textcomp}
\usepackage[cache=true, cachedir=minted-cache, frozencache=true]{minted}

\usepackage[percent]{overpic}
\usepackage[list=true,listformat=simple,margin=10pt]{subcaption}
\usepackage{hepnames}

\journal{Nucl. Instr. Meth. A}

\newcommand{\apsq}{\texorpdfstring{\ensuremath{\mathrm{Allpix}^2}}{Allpix\textasciicircum 2}\xspace}

\begin{document}
\begin{frontmatter}
  \title{Transient Monte Carlo Simulations for the Optimisation and Characterisation of Monolithic Silicon Sensors}

\author[cern]{R.~Ballabriga}
\author[cern]{J.~Braach\fnref{uhamburg}}
\author[cern]{E.~Buschmann}
\author[cern]{M.~Campbell}
\author[cern]{D.~Dannheim}
\author[cern]{K.~Dort\corref{corr}\fnref{ugiessen}}
\ead{katharina.dort@cern.ch}
\author[desy]{L. Huth}
\author[cern]{I.~Kremastiotis}
\author[cern]{J.~Kr\"oger\fnref{heidelberg}}
\author[cern]{L.~Linssen}
\author[ugeneva]{M.~Munker}
\author[desy]{P.~Sch\"utze}
\author[cern]{W.~Snoeys}
\author[desy]{S.~Spannagel}
\author[desy]{T.~Vanat}

\address[cern]{CERN, Geneva, Switzerland}
\address[desy]{Deutsches Elektronen-Synchrotron DESY, Notkestr. 85, 22607 Hamburg, Germany}
\address[ugeneva]{University of Geneva, Geneva, Switzerland}

\cortext[corr]{Corresponding author}

\fntext[uhamburg]{Also at University of Hamburg, Germany}
\fntext[ugiessen]{Also at University of Giessen, Germany}
\fntext[heidelberg]{Also at University of Heidelberg, Germany}

\begin{abstract}

	An ever-increasing demand for high-performance silicon sensors requires complex sensor designs that are challenging to simulate and model.
	The combination of electrostatic finite element simulations with a transient Monte Carlo approach provides simultaneous access to precise sensor modelling and high statistics.
	The high simulation statistics enable the inclusion of Landau fluctuations and production of secondary particles, which offers a realistic simulation scenario.
	The transient simulation approach is an important tool to achieve an accurate time-resolved description of the  sensor, which is crucial in the face of novel detector prototypes with increasingly precise timing capabilities.
	The simulated time resolution as a function of operating parameters as well as the full transient pulse can be monitored and assessed, which offers a new perspective for the optimisation and characterisation of silicon sensors.

	In this paper, a combination of electrostatic finite-element simulations using 3D TCAD and transient Monte Carlo simulations with the \apsq framework are presented for a monolithic CMOS pixel sensor with a small collection electrode, that is characterised by a highly inhomogeneous, complex electric field.
	The results are compared to transient 3D TCAD simulations that offer a precise simulation of the transient behaviour but long computation times.
	Additionally, the simulations are benchmarked against test-beam data and good agreement is found for the performance parameters over a wide range of different operation conditions.
\end{abstract}


\end{frontmatter}

\tableofcontents

\section{Introduction}
\label{sec:introduction}
\input{introduction}

\section{Sensor Design}
\label{sec:clictd}

\input{clictd_sensor}


\section{TCAD Simulation}
\label{sec:tcad}
 \input{tcad_simulations}

\section{Transient Monte Carlo Simulation}
\label{sec:mc_transient}
 \input{mc_transient}


\section{Evaluation of Transient Sensor Response}
\label{sec:tcad_mc_comparison}
\input{tcad_mc_comparison}
\label{sec:time_resolution}
\input{time_resolution}

\section{Reconstruction and Analysis }
\label{sec:reco_analysis}
\input{reco_analysis}
\input{uncertainties}

\section{Comparison with Test-Beam Data}
\label{sec:data_comparison}
\input{data_comparison}

\section{Summary \& Outlook}
\label{sec:summary}
\input{summary}

\section*{Acknowledgements}
\label{sec:acknowledgements}
\input{acknowledgements}

\section*{CRediT authorship statement}
\label{sec:credit}
\input{credit_statement}

\bibliography{bibliography}

\end{document}

%% file: introduction.tex
 The simulation of silicon detectors is an important ingredient in the development and characterisation of novel prototypes for future particle physics experiments and other applications.
Simulation is key in the design and performance optimisation, in the interpretation of measurement results and in the comprehension of underlying mechanisms.
It is also an aid in the optimisation of analysis and reconstruction algorithms and is therefore an indispensable tool in various stages in the detector development cycle.

The combination of 3-dimensional finite-element simulations using Technology Computer-Aided Design (3D TCAD) with the Monte Carlo framework \apsq~\cite{apsq} was validated for a monolithic CMOS sensor with a small collection electrode and it was shown that the advantages of both simulation approaches can be exploited simultaneously: detailed sensor modelling and high simulation rates that enable the inclusion of statistical fluctuations~\cite{allpix-hrcmos}.
The \apsq framework allows for an end-to-end simulation of the response of a silicon sensor to the traversal of highly energetic particles, from initial energy deposition to signal digitisation.

In light of new detector prototype developments with sub-nanosecond time resolutions, time-resolved simulations are of utmost importance.
Transient Monte Carlo simulations are provided by the \apsq framework on the basis of the Shockley-Ramo theorem~\cite{shockley, ramo}.
This allows the investigation of the current pulse induced on the sensor electrodes and pave the way for the optimisation of sensor designs with fast signal formation.

In this document, the transient Monte Carlo simulations are validated against transient 3D TCAD simulations and test-beam data from a monolithic silicon sensor prototype with a small collection electrode.
The device is characterised by complex field configurations as well as doping concentrations that range over several orders of magnitude.
A precise modelling of the sensor is therefore crucial to make accurate predictions about its performance.


%% file: clictd_sensor.tex


\begin{figure*}[bhtp]
	\centering
	\begin{subfigure}[t]{0.33\textwidth}
		\centering
		\includegraphics[width=1.2\columnwidth]{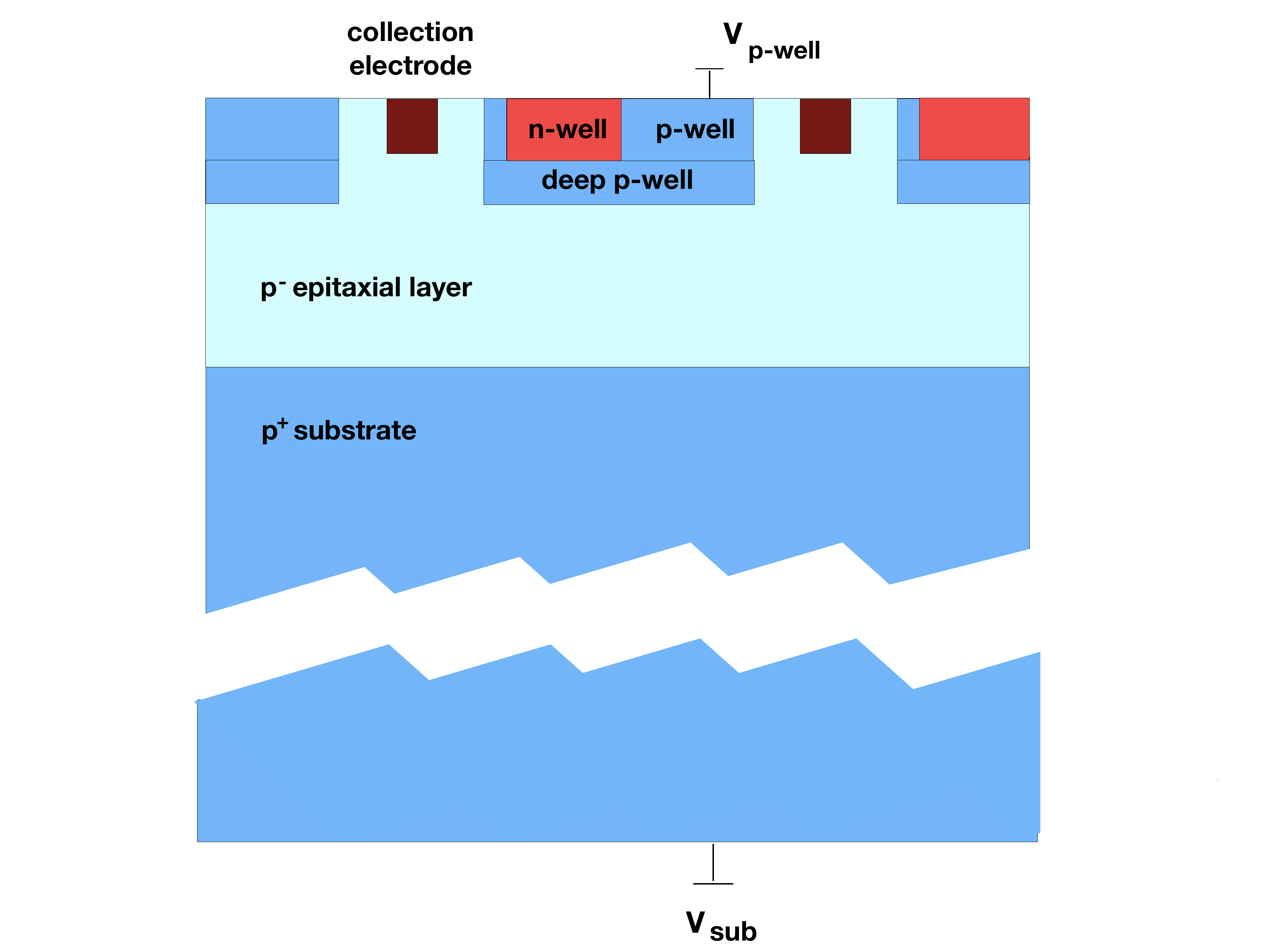}
		\caption{No n-type implant}
		\label{fig:clictdSensor:left}
	\end{subfigure}%
	\begin{subfigure}[t]{0.33\textwidth}
		\centering
		\includegraphics[width=1.2\columnwidth]{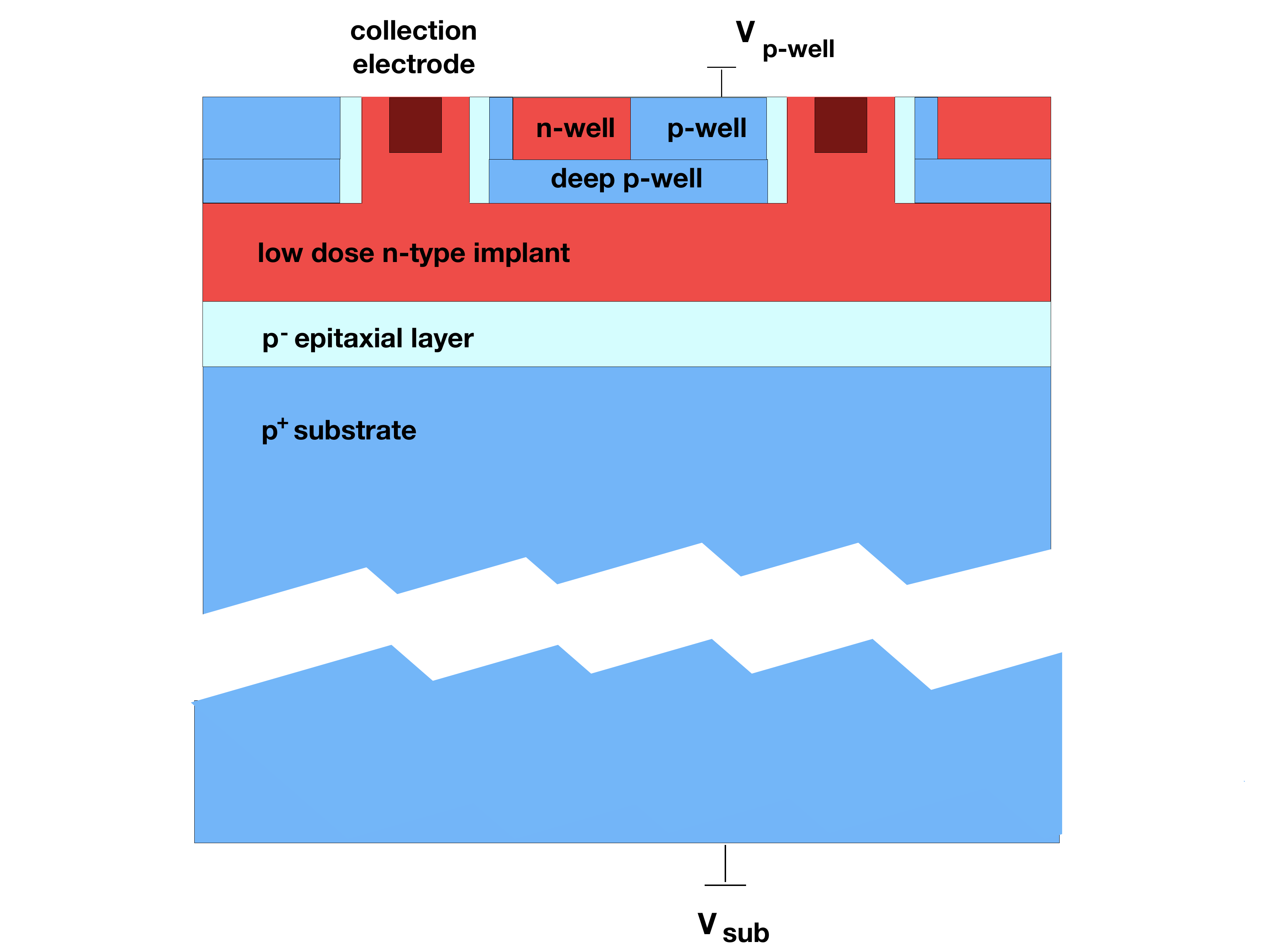}
		\caption{Continuous n-type implant}
		\label{fig:clictdSensor:center}
	\end{subfigure}%
	\begin{subfigure}[t]{0.33\textwidth}
		\centering
		\includegraphics[width=1.2\columnwidth]{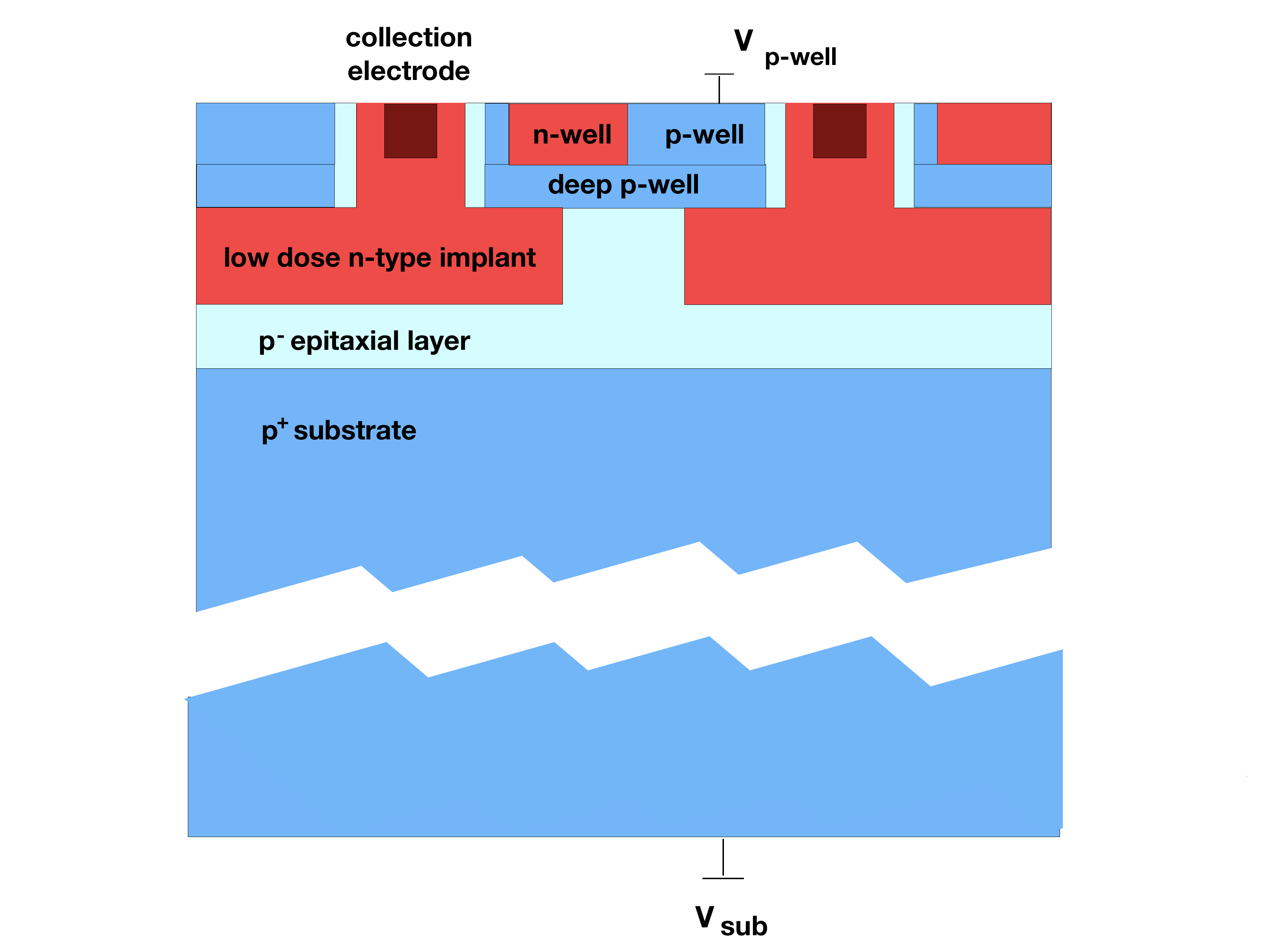}
		\caption{Segmented n-type implant}
		\label{fig:clictdSensor:right}
	\end{subfigure}%
\caption{Schematic representation of the three different pixel flavours investigated. They consist of a p$^+$ substrate, a p$^-$ epitaxial layer and different well layouts with no~(\subref{fig:clictdSensor:left}), with continuous~(\subref{fig:clictdSensor:center}) and with segmented n-implant~(\subref{fig:clictdSensor:right}) (not to scale).}
	\label{fig:sensor_designs}
\end{figure*}

The sensors investigated here are implemented in a 180\,nm CMOS imaging process, as shown schematically in Fig.~\ref{fig:sensor_designs}~\cite{tj-modified}.
A single pixel cell is defined as the rectangular region centred around each collection electrode, with edges equidistant between neighbouring electrodes. 
Therefore, the figure displays two adjacent pixels for each flavour. 
They exhibit a small n-type collection electrode placed on top of a \SI{30}{\micro m} thick high-resistivity epitaxial layer grown on top of a p-type substrate.
In this paper, prototypes with a total thickness of \SI{100}{\micro m} are investigated.
The pixel readout electronics are placed in deep p-wells.
Bias voltages of $\textrm{V}_\textrm{sub}$ and $\textrm{V}_\textrm{pwell}$ are applied to the substrate and nodes in the deep p-wells, respectively.
In this work, the sensors are operated with a reverse bias of  $\textrm{V}_\textrm{sub} = \textrm{V}_\textrm{pwell} = -6$\,V.
While full lateral depletion is achieved at these bias voltages, the depleted volume only extends to $21 - \SI{23}{\micro \meter}$ in depth.

The pixels come in three flavours, differentiated by the option of a deep low-dose n-implant. 
The first flavour has no such implant as shown in Fig.~\ref{fig:clictdSensor:left}.
The second flavour has a continuous deep n-implant that ensures full lateral depletion of the epitaxial layer~\cite{tj-modified} as indicated in Fig.~\ref{fig:clictdSensor:center}.
In the third pixel flavour, the deep n-implant is segmented as shown in Fig.~\ref{fig:clictdSensor:right}, which generates a lateral electric field leading to a faster charge collection~\cite{Munker:2019vdo}.

While the comparison of the transient current pulse with transient 3D TCAD will be performed for all three pixel flavours, the comparison with test-beam data is only shown for the flavour with continuous n-implant, using a \SI{100}{\micro m} thick sample.
The comparison for the flavour with no n-implant and the one with segmented n-implant produce similar results and can be found elsewhere~\cite{allpix-hrcmos, Dort:2773808}.

The continuous n-implant design is used for the CLIC Tracker Detector (CLICTD) sensor, a technology demonstrator developed in the context of the tracking detector studies for the Compact Linear Collider (CLIC)~\cite{clictd_design_characterization}.

The CLICTD sensor has an active matrix of 16 columns $\times$ 128 rows (\textit{detection channels}), each measuring  \SI{300x30}{\micro m}.
The detection channels are segmented into eight sub-pixels along the \SI{300}{\micro m} dimension, yielding a sub-pixel pitch of \SI{37.5x30}{\micro m}.
Each sub-pixel is equipped with its own collection electrode and analogue front-end.
The output of the eight sub-pixels are combined by an \textit{OR} gate in a common digital front-end of the detection channel, allowing for a reduction of the digital logic.
The CLICTD sensor features simultaneous 8-bit Time-of-Arrival (ToA) (\SI{10}{ns} ToA bins) and 5-bit Time-over-Threshold (ToT) measurement capabilities, although these are not studied in this paper (see~\cite{clictdTestbeam} for further details).

%% file: tcad_simulations.tex
Three-dimensional TCAD simulations using the Synopsys Sentaurus framework~\cite{sentaurus-website} are used to model the electrostatic properties of the sensor, with the geometry as in Section~\ref{sec:clictd}.
The TCAD simulation comprises only the epitaxial layer of the sensor.
A single pixel cell is simulated and periodic boundary conditions are applied.
The mesh is adjusted to the doping gradient in the structure i.e. a finer mesh is applied in regions with a high gradient such as the p-wells or around the implants.


\subsection{Electrostatic Simulation}

\begin{figure}[tbp]
	\centering
	\includegraphics[width=.8\columnwidth]{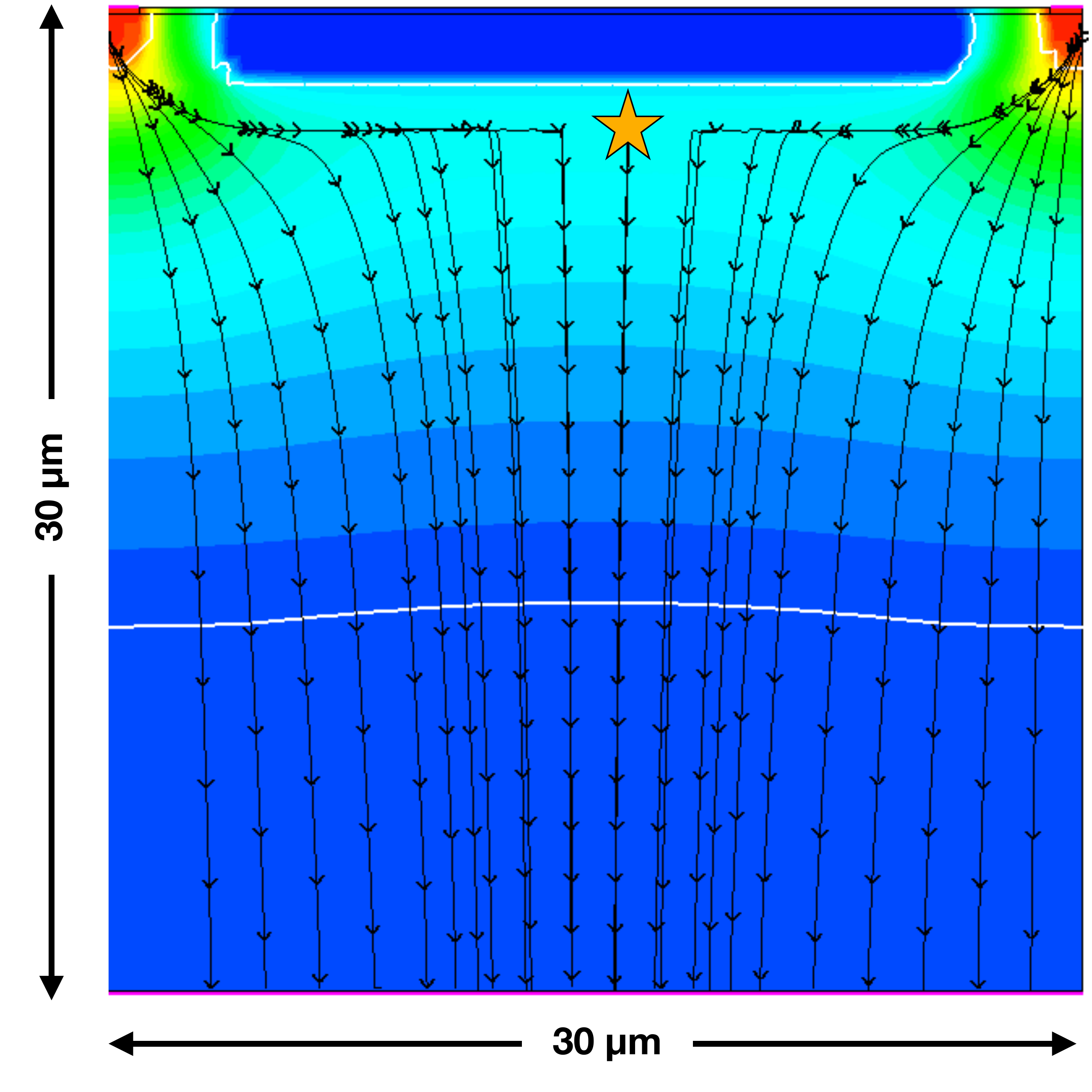}%
	\caption{Cross section through the implants of a 3D TCAD simulation showing the electrostatic potential and electric field streamlines in the epitaxial layer of the pixel flavour with continuous n-implant.
	The white line indicates the edge of the depleted volume and the star denotes the electric field minimum.
	The collection electrodes are located at the top-left and top-right edge of the structure.}
	\label{fig:ePotentialTCAD}
\end{figure}

A cross section of the electrostatic potential of the pixel flavour with the continuous n-implant is depicted in Fig.~\ref{fig:ePotentialTCAD}.
The complexity and non-linearity of the potential is a direct consequence of the small collection electrode design and demands precise modelling of the sensor for accurate performance predictions.
The black arrows denote the electric field streamlines and the star indicates an electric field minimum below the p-wells at the pixel edges.
Charge carriers close to these edges first drift into the field minimum before they propagate to the collection electrode, resulting in a slow charge collection and increased charge sharing, as detailed in~\cite{Munker:2019vdo}.
As a consequence, the impact of charge carrier diffusion and recombination are crucial in this part of the pixel cell.

The white line at the bottom of the structure shows the border of the depleted volume indicating that a small part at the bottom of the epitaxial layer is not fully depleted.
Moreover, the depleted volume does not reach into the substrate.
In these regions, the impact of diffusion is particularly high.

3D TCAD calculations are used to provide a precise sensor description for the \apsq simulation.
For the modelling of the CLICTD sensor, the electromagnetic field, the weighting potential and the doping profile are included.
The maps are converted to a regularly spaced mesh using the \apsq~\textit{Mesh Converter} tool for a faster and less computing-intense lookup of electric field vectors at different positions in the sensor.

\subsection{Transient Simulation}

Transient 3D TCAD simulations are used to study the charge collection behaviour of the sensor.
In the simulation, 63 electron-hole pairs per micrometer~\cite{Meroli:2011zzb} are injected along a straight line at the pixel corners perpendicular to the sensor surface.
The injection line is smeared laterally by a Gaussian distribution with a standard deviation of \SI{50}{nm} and a finer mesh around the injection line is used.

The drift and diffusion equations are numerically solved for each predefined time step and the time-resolved response of the sensor is extracted.
The step size is adapted to the simulated sensor and to the current simulation time i.e. finer time steps are used right after the charge injections and coarse steps when most of the charge carriers have recombined or reached the electrode.

Transient 3D TCAD simulations provide a detailed and self-consistent solution to the electromagnetic response of the sensor to charge injection.
However, the high computation times render these simulations unpractical for performance studies where large simulation samples are needed.

%% file: mc_transient.tex
The Monte Carlo simulations performed in this paper largely follow the setup outlined in~\cite{allpix-hrcmos}, with a few key differences with respect to energy deposition, charge carrier transport and signal formation.
They were performed with \apsq version 2.0 and profited significantly from the multithreading capabilities introduced with this version.

\subsection{Energy Deposition}

For the initial deposition of energy in the sensor by the ionising radiation, two models have been used.
The first model seeks to replicate the simplified situation simulated in the transient TCAD calculations presented in the previous section, to allow for a direct comparison between the two methods.
For this, the \apsq framework provides the \emph{DepositionPointCharge} module which allows the generation of electron-hole pairs at a defined position or along a line through the sensor as Linear Energy Transfer (LET).
The respective configuration is shown in Listing~\ref{lst:depositionpointcharge}.
Here, the position of the charge carrier deposition was chosen as the corner between two diagonally adjacent pixels (10,10) and (11,11) of the sensor in accordance with the position of the energy deposition simulated with transient TCAD, by setting the \texttt{position} parameter to the desired location in local coordinates.
The parameters \texttt{number\_of\_steps} and \texttt{number\_of\_charges} define the distribution of deposited charge carriers in the sensor by specifying the total number of steps along the path, in which charge is deposited,  and the equivalent number of charge carriers deposited per micrometre.

\begin{listing}
\begin{minted}[frame=single,framesep=3pt,breaklines=true,tabsize=2]{ini}
[DepositionPointCharge]
source_type = "mip"
model = "fixed"
position = 315um, 393.75um
number_of_steps = 30
number_of_charges = 63/um
\end{minted}
\caption{Configuration section for the \emph{DepositionPointCharge} module used to replicate the simplified energy deposition and charge carrier generation used in the transient 3D TCAD simulations. Here, a model with fixed position and a MIP-like LET is used.}
\label{lst:depositionpointcharge}
\end{listing}

For the comparison with test-beam data presented in this paper, the \emph{DepositionGeant4} module is used.
The module uses Geant4~\cite{geant4, geant4-2} to simulate the creation of secondary particles and a more realistic interaction of the incoming particle with the detector material.
The corresponding configuration can be obtained from~\cite{allpix-hrcmos}.

\subsection{Shockley-Ramo Theorem \& Weighting Potential}

The calculation of the induced current of a moving charge carrier requires the knowledge of the weighting potential in addition to the electric field of the sensor.
The weighting potential for a given sensor geometry can be calculated analytically or by means of a finite-element simulation in TCAD by setting the electrode of the pixel under consideration to unit potential, and all other electrodes to ground~\cite{planecondenser}.

The Shockley-Ramo theorem~\cite{shockley, ramo} then states that the charge $Q_n^{ind}$ induced by the motion of a charge carrier is equivalent to the difference in weighting potential between the previous location $\vec{x}_0$ and its current position $\vec{x}_1$, viz.\
\begin{equation}
Q_n^{ind}  = \int_{t_0}^{t_1} I_n^{ind} \textrm{d}t = q \left[ \phi (\vec{x}_1) - \phi(\vec{x}_0) \right],
\label{eq:shockleyramo}
\end{equation}
assuming discrete time steps, as detailed in Section~\ref{sec:signalFormation}.
Here, $q$ is the charge of the carrier, $\phi(\vec{x})$ the weighting potential at position $\vec{x}$ and $ I_n^{ind}$ the induced current in the particular time step.

In \apsq, weighting potentials can be loaded and applied to individual sensors.
Depending on the geometry of the sensor and the pixel cell, it might be necessary to calculate a weighting potential for a matrix of \num{3x3} pixels to cover induction in neighbouring pixels, or just for a single pixel if the weighting potential is confined to the volume very close to the collection electrode.
Here, the weighting potential of a single pixel is considered, owing to the small ratio between the size of the collection electrode and active sensor thickness.

\subsection{Charge Carrier Lifetime \& Recombination}

The recombination of charge carriers within the silicon lattice needs to be taken into account when simulating the signal formation, especially with highly-doped regions present in the sensor, such as the substrate wafer of CLICTD.

A charge carrier is given a probability of
\begin{equation}
    \label{eq:recomb:prob}
    p = e^{- \Delta t / \tau(N)}
\end{equation}
of surviving a simulation time step $\Delta t$ without recombining with the lattice, with $\tau(N)$ being the lifetime calculated from the local doping concentration as described in the following.
Two main recombination mechanisms are relevant for the silicon sensors investigated in this paper: the Shockley-Read-Hall and the Auger recombination.

\paragraph{Shockley-Read-Hall recombination} The recombination process can be induced by defects and impurities which create additional energy levels in the band gap as described by the Shockley-Read-Hall (SRH) model~\cite{fossum-lee, fossum}.
If the recombination centres are close to the middle of the band gap, the charge carrier lifetime $\tau_\mathrm{SRH}$ is given by:
\begin{equation}
	\tau_\mathrm{srh}(N_{d,e/h}) = \frac{\tau_{0, e/h}}{1 + N_{d, e/h}/N_{d0, e/h}}
\end{equation}

where $N_{d, e/h}$ is the doping concentration. $\tau_{0, e/h}$ and $N_{d0, e/h}$ are the reference lifetime and reference doping concentration, respectively, which are taken from~\cite{fossum-lee} as
\begin{equation*}
    \begin{split}
        \tau_{0,e} &= \SI{1e-5}{s} \\
        N_{d0,e}   &= \SI{1e16}{\per \cubic \cm} \\
    \end{split}
    \qquad
    \begin{split}
        \tau_{0,h} &= \SI{4.0e-4}{s} \\
        N_{d0,h}   &= \SI{7.1e15}{\per \cubic \cm} \\
    \end{split}
\end{equation*}
for electrons and holes, respectively.

\paragraph{Auger recombination} This recombination mechanism becomes increasingly important at high doping levels exceeding $\SI{5e18}{\centi\metre^{-3}}$~\cite{FOSSUM1983569}.
The model assumes that the excess energy created by electron-hole recombinations is transferred to another electron (\textit{e-e-h process}) or another hole (\textit{e-h-h process}).
The total recombination rate is then given by~\cite{augerKerr2002}:
\begin{equation}
	R_{a} = C_n n^2p + C_p n p^2
\end{equation}
where $C_n$/$C_p$ are the Auger coefficients and $n$/$p$ the free charge carrier concentrations.
The first term corresponds to the e-e-h process and the second term to the e-h-h process.
In highly-doped silicon, the Auger lifetime for minority charge carriers can be written as
\begin{equation}
	\tau_\mathrm{a} = \frac{1}{C_a \cdot N_d^2}
\end{equation}
where $C_a = C_n + C_p$ is the ambipolar Auger coefficient and $N_d$ is the majority carrier density.
The coefficients can be determined experimentally and for the simulations presented in this paper, the values have been taken from~\cite{DziewiorSchmid} as
\begin{equation*}
    \begin{split}
        C_n &= \SI{2.8e-31}{\centi\metre^{-6}\per\second} \\
        C_p &= \SI{0.99e-31}{\centi\metre^{-6}\per\second},
    \end{split}
\end{equation*}
resulting in $C_a = \SI{3.79e-31}{\centi\metre^{-6}\per\second}$.


\paragraph{Combination} The two charge carrier recombination models are combined by inversely summing the individual lifetimes calculated by the models via
\begin{equation}
    \label{eq:recom:combined}
    \tau^{-1}(N_d) =  \left\{
    \begin{array}{ll}
        \tau_{srh}^{-1}(N_d) + \tau_{a}^{-1}(N_d) & \textrm{(\emph{minority})} \\
        \tau_{srh}^{-1}(N_d) & \textrm{(\emph{majority})}
    \end{array}
    \right.
\end{equation}
where the Auger lifetime is only taken into account for minority charge carriers.

\subsection{Charge Carrier Mobility}

A combination of the low-field Masetti~\cite{masetti} and Canali~\cite{canali} mobility models for charge carrier mobility in silicon is used.
This allows the dependence on both the electric field and the doping concentration to be taken into account.
In particular, the saturation velocity is considered, which is crucial to avoid an overestimation of the charge carrier mobility. 
The mobility is parametrised as a function of the electric field strength $E$ and the doping concentration $N$:
\begin{equation}
    \label{eq:mob:mascan}
    \mu (E, N) = \frac{\mu_{m}(N)}{\left(1 + \left(\mu_{m}(N) \cdot E / v_{m} \right)^{\beta} \right)^{1 / \beta}}
\end{equation}
where $\mu_{m}(N)$ is the Masetti mobility; $v_m$ and $\beta$ are phenomenological parameters from the Canali model, defined for electrons and holes separately.

A more detailed description of the individual models and their combination can be found in the \apsq User Manual~\cite{apsq_manual}.

\subsection{Signal Formation}
\label{sec:signalFormation}

\begin{figure*}[htbp]
  \centering
  \begin{subfigure}[t]{0.33\textwidth}
    \centering
    \includegraphics[width=0.7\columnwidth]{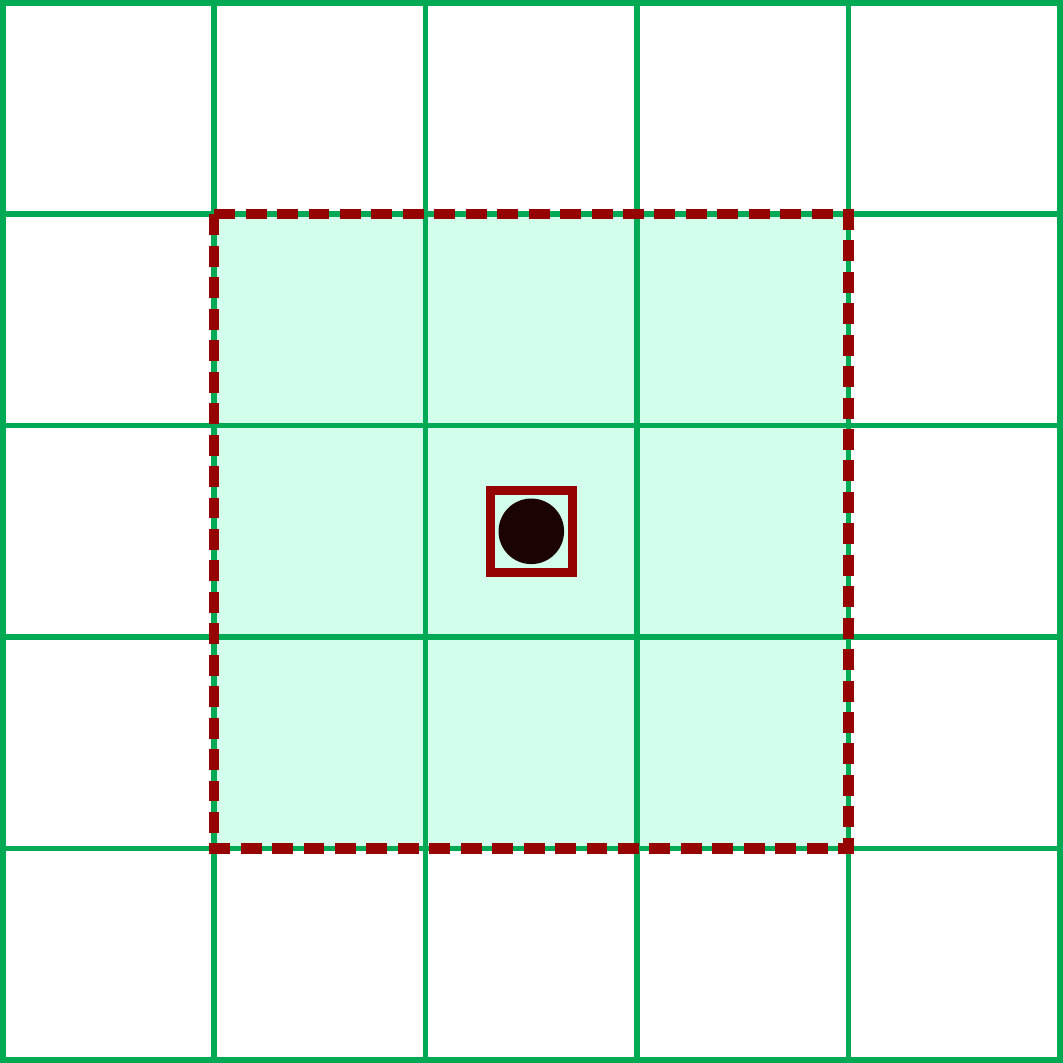}
    \label{fig:induction:center}
    \caption{Induced current on center pixel}
    \end{subfigure}%
  \begin{subfigure}[t]{0.33\textwidth}
    \centering
    \includegraphics[width=0.7\columnwidth]{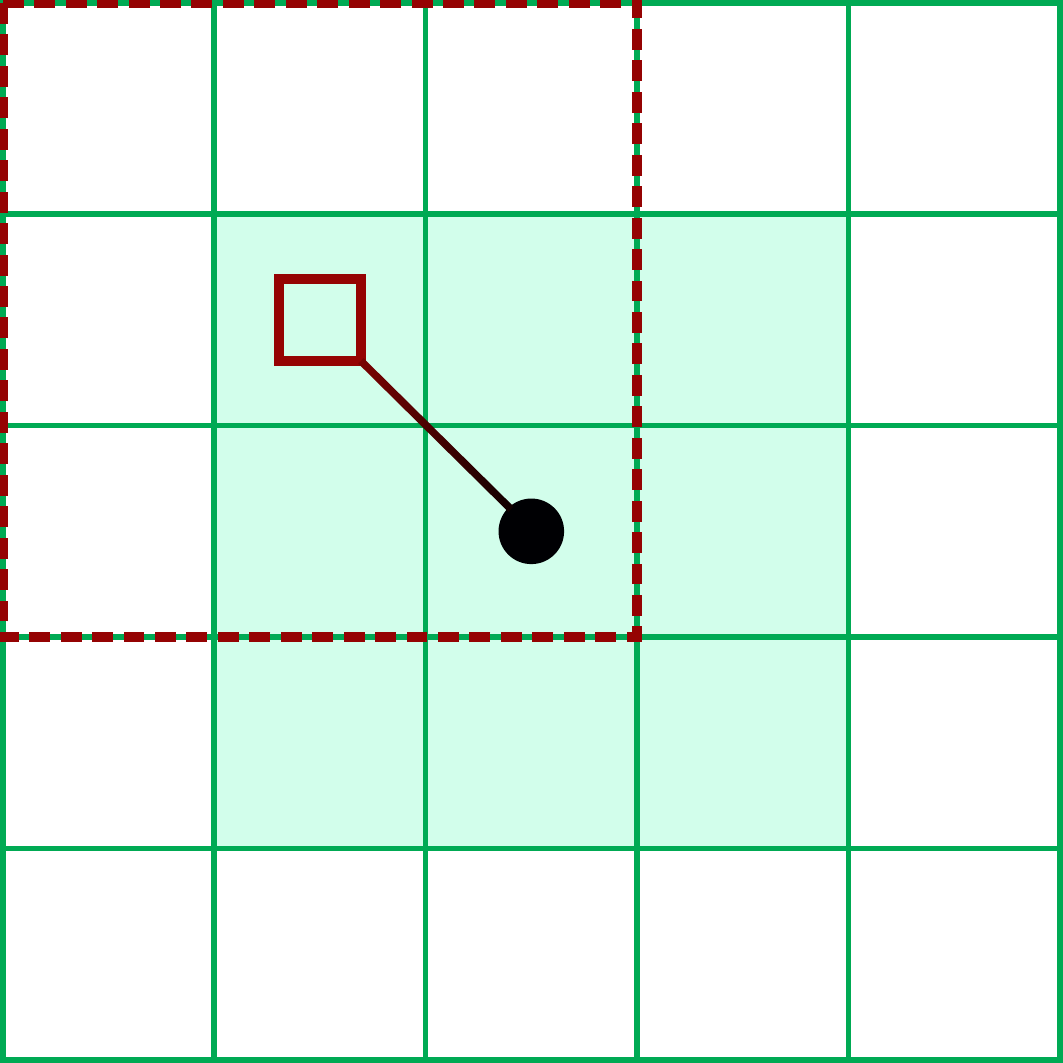}
    \label{fig:induction:topleft}
    \caption{Induced current on upper-left pixel}
  \end{subfigure}%
  \begin{subfigure}[t]{0.33\textwidth}
    \centering
    \includegraphics[width=0.7\columnwidth]{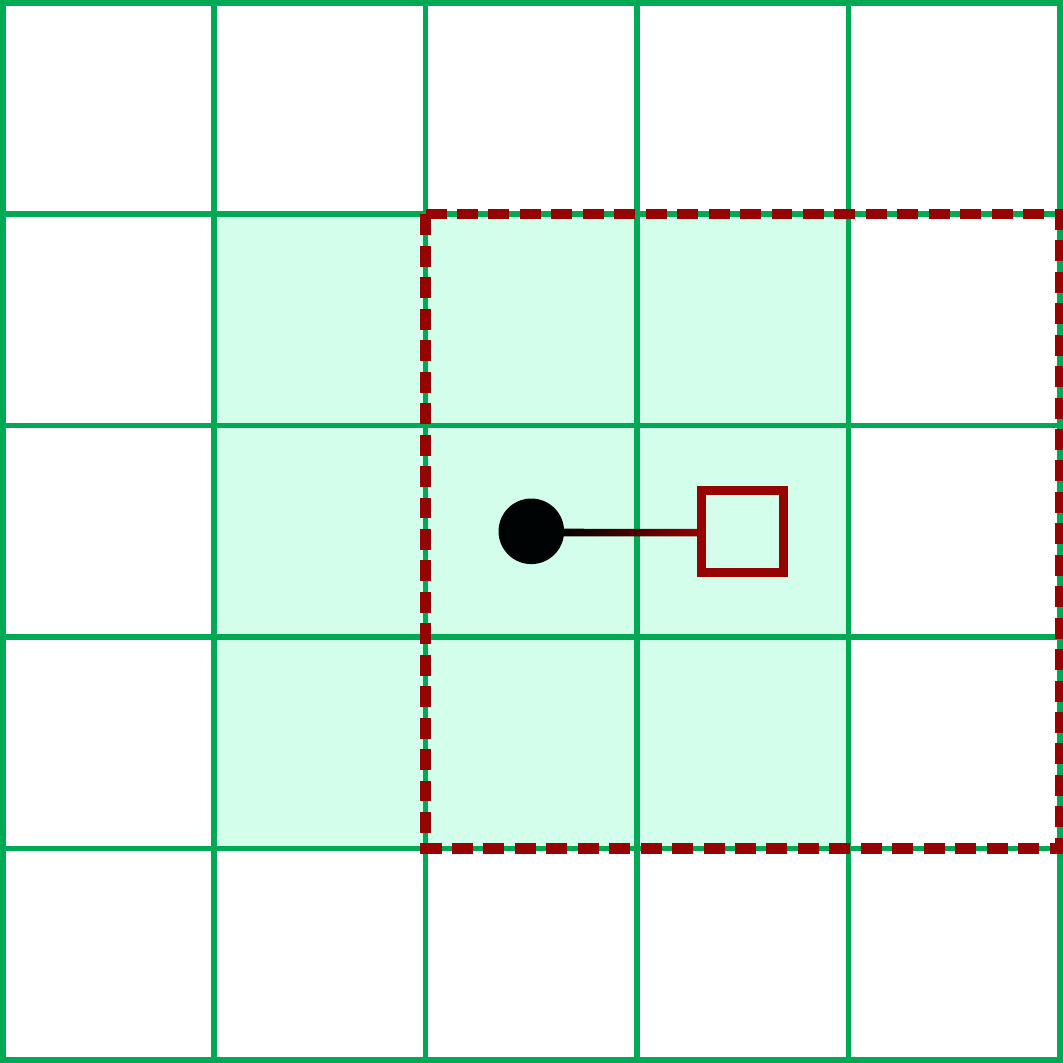}
    \label{fig:induction:centerright}
    \caption{Induced current on center-right pixel}
    \end{subfigure}%
  \caption{Shifting of a \num{3x3} weighting potential over a \num{3x3} region of interest. Here, the pixels in the region of interest for which the induced current is calculated are shown in green. The charge carrier position is indicated by the black dot and the weighting potential is displayed in red, with its electrode at unit potential as small square and its full extent indicated by the dashed line.}
  \label{fig:induction}
\end{figure*}

\begin{listing}
\begin{minted}[frame=single,framesep=3pt,breaklines=true,tabsize=2]{ini}
[TransientPropagation]
temperature = 293K
charge_per_step = 10
timestep = 7ps
distance = 1
integration_time = 50ns
mobility_model = "masetti_canali"
recombination_model = "srh_auger"
\end{minted}
\caption{Configuration section for the \emph{TransientPropagation} module used to simulate the charge transport.}
\label{lst:transientpropagation}
\end{listing}

For the simulation of charge carrier motion, the \emph{TransientPropagation} module is used.
Similar to the \emph{GenericPropagation} module used in~\cite{allpix-hrcmos}, it employs the Runge-Kutta-Fehlberg (RKF) integration method~\cite{fehlberg} but, in contrast, uses fixed time steps to simplify the assignment to time bins in the generated current pulse.

The configuration used for the charge carrier transport is provided in Listing~\ref{lst:transientpropagation}.
Here, the parameters \texttt{charge\_per\_step} and \texttt{timestep} control how many charge carriers are moved together and how fine the timesteps of the propagation are, respectively.
The \texttt{distance} parameter allows the configuration of the maximum distance of neighbouring pixels for which the induced currents are calculated.
For a value of $0$, only the current for the central pixel is calculated, below which the charge carrier is located.
A distance of $1$ includes all direct neighbours.
The total integration time is set to \SI{50}{ns} for the simulation that is compared to test-beam data.
The value  corresponds to the estimated integration time of the CLICTD front-end and is significantly larger than the expected signal formation time for the pixel flavour with continuous n-implant.

For each of the charge carrier groups propagated together, the algorithm repeats the following steps until a sensor surface has been reached or the charge carriers have recombined:
\begin{enumerate}
    \item The charge carrier mobility and resulting velocity are calculated at the current position.
    \item A step is performed by the RKF integration method.
    \item The offset caused by diffusion is calculated and added to the new position.
    \item Determine whether the charge carrier has recombined with the silicon lattice via Eqs.~\ref{eq:recomb:prob}~and~\ref{eq:recom:combined}, or the charge carrier has left the sensor volume.
    \item The closest pixel is determined.
    \item For the closest and each of the surrounding pixels, the current induced by the step is calculated via Eq.~\ref{eq:shockleyramo}.
\end{enumerate}

The calculation of the induced current in the different pixels is demonstrated in Figure~\ref{fig:induction}.
The weighting potential is centred with its readout electrode on unit potential on the pixel of interest, for which the induced current by the charge carrier movement is to be calculated.
For the subsequent pixel of interest, the position of the weighting potential is adjusted accordingly.
For the special case of a strongly confined weighting potential at the collection electrode, it suffices to consider the  potential of a single pixel cell.

\subsection{Current Pulses \& Digitisation}

The final per-pixel current pulses are formed from the currents induced by individual charge carriers in the \emph{PulseTransfer} module.
The resulting pulses can either be analysed directly or further processed by a front-end simulation module such as the \emph{DefaultDigitizer}, or the \emph{CSADigitizer} to calculate derived quantities such as the ToA or ToT.
For the comparison with data, the \emph{DefaultDigitizer} was used.

%% file: tcad_mc_comparison.tex
\begin{figure*}[bpt]
	\centering
	\begin{subfigure}[t]{0.32\textwidth}
		\centering
		\includegraphics[width=\columnwidth]{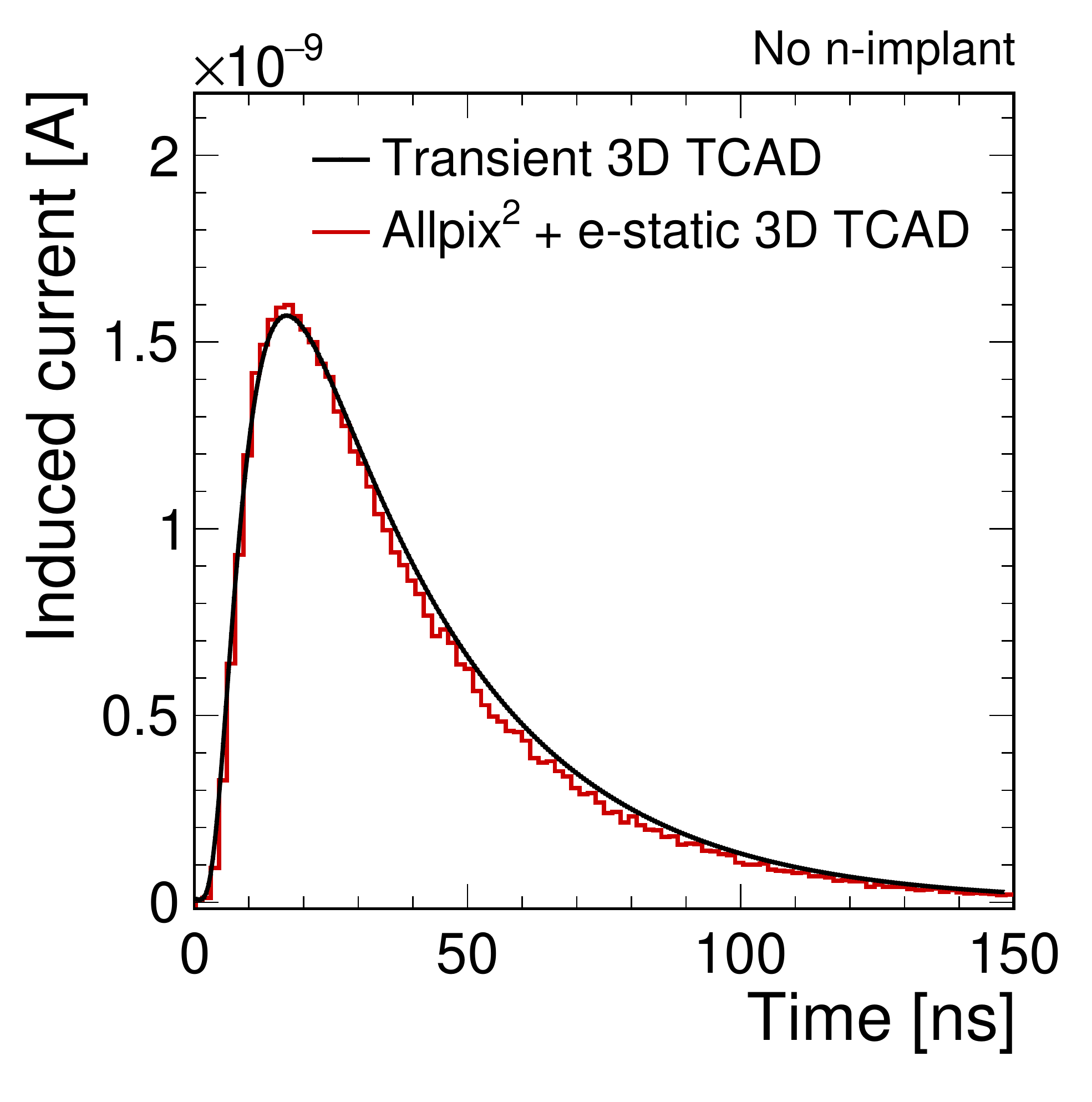}
		\label{fig:transientComparison:center}
		\caption{}
	\end{subfigure}%
	\begin{subfigure}[t]{0.32\textwidth}
		\centering
		\includegraphics[width=\columnwidth]{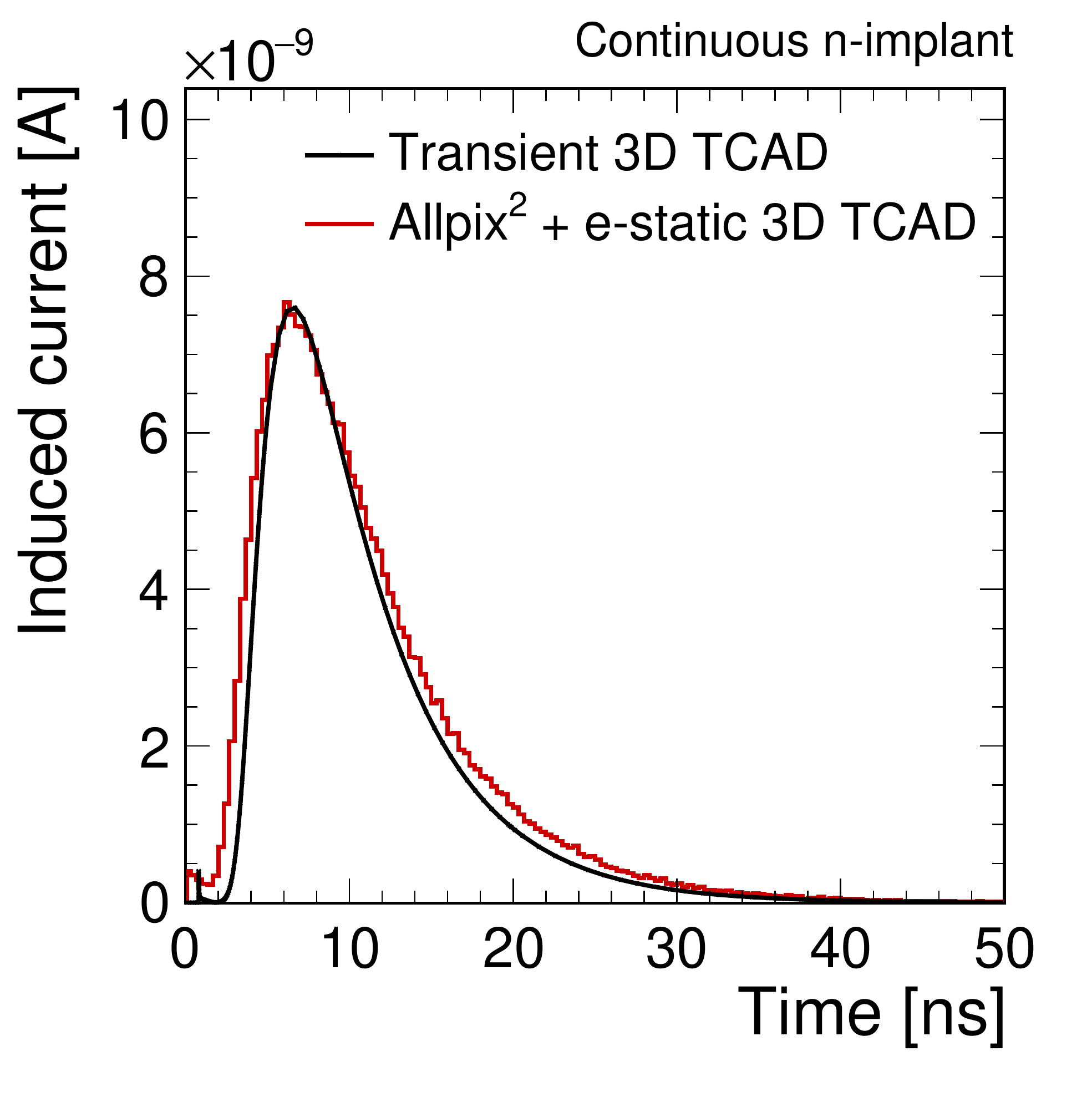}
		\label{fig:transientComparison:left}
		\caption{}
	\end{subfigure}%
	\begin{subfigure}[t]{0.32\textwidth}
		\centering
		\includegraphics[width=\columnwidth]{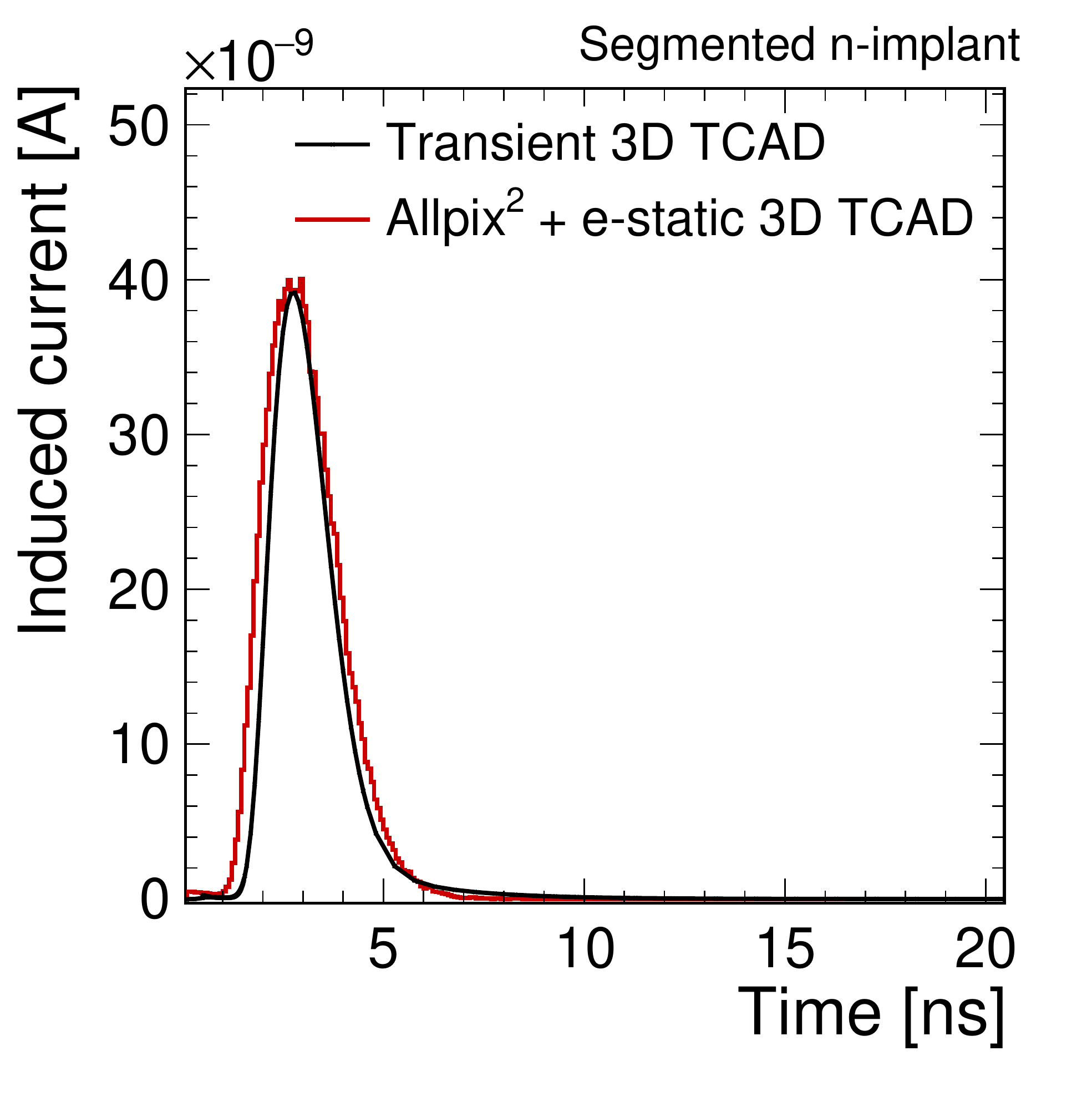}
		\label{fig:transientComparison:right}
		\caption{}
	\end{subfigure}%
	\caption{Transient current pulses for transient 3D TCAD and \apsq combined with electrostatic 3D TCAD after charge injection at the pixel corner.
	Three different sensor designs were tested: a design with (a) no n-implant, (b) continuous n-implant, (c) segmented n-implant for the CLICTD geometry using a pixel pitch of \SI{37.5x30}{\micro m} and an epitaxial layer thickness of \SI{30}{\micro m}. 
	All simulations are performed using the same bias voltage of $\textrm{V}_\textrm{sub} = \textrm{V}_\textrm{pwell} = -6$\,V.}
	\label{fig:transientComparison}
\end{figure*}

For the small collection electrode design, the optimisation of the sensor properties is especially important at the pixel corners, since the distance to the collection electrode is maximal and the charge collection therefore slower compared to the rest of the pixel cell.
It also represents a challenging simulation scenario since charge carriers created at the pixel corners traverse a large part of the active sensor volume, thereby probing the field distribution inside the pixel cell.

\subsection{Comparison with TCAD}
In transient 3D TCAD, the pixel corner is investigated by injecting electron-hole pairs along a straight line at the intersection of four neighbouring pixels, as explained in Section~\ref{sec:tcad}.
The same simulation conditions are replicated with \apsq using the static field maps obtained from the electrostatic 3D TCAD simulation.
The average pulse of 50 \apsq simulation events is shown to smooth out statistical fluctuations, except for the design without n-implant, where 200 simulation events are averaged.
The simulation time per event is about 0.1 - 0.2\,s compared to typically 8\,h using 3D TCAD on the same machine and the same number of threads.

The resulting current pulses induced on one pixel cell are shown in Fig.~\ref{fig:transientComparison} for transient 3D TCAD and \apsq combined with electrostatic 3D TCAD.
The pulses for all three sensor designs are displayed.

\begin{figure}[tbp]
	\centering
	\includegraphics[width=\columnwidth]{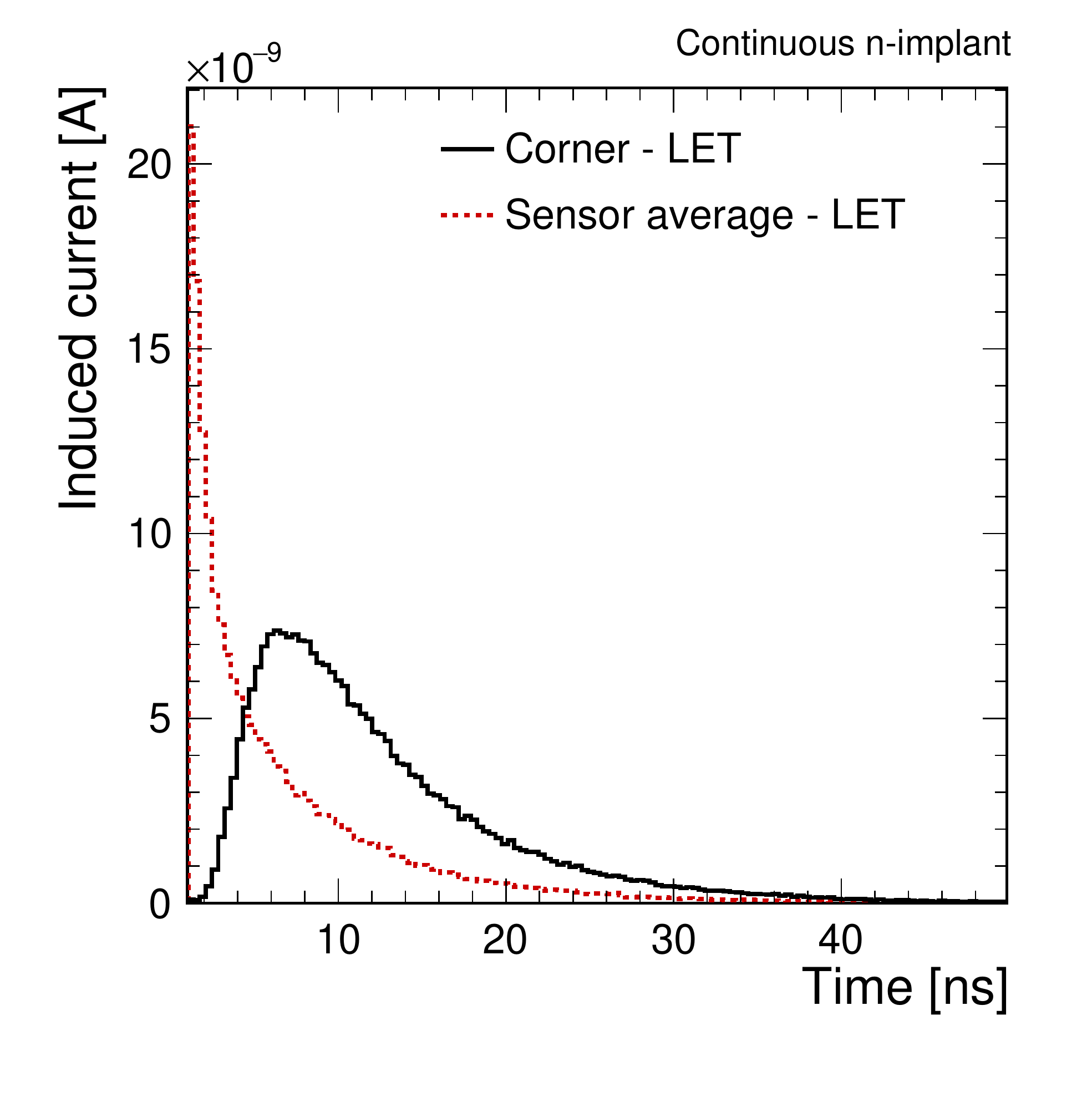}%
	\caption{Induced current for charge carriers homogeneously injected across the pixel cell and solely in the pixel corner.}
	\label{fig:transientPulseModCornerAllSensor}
\end{figure}

The different pixel flavours have significant differences in charge collection times and field configurations, but it can be seen that the two different simulation approaches give compatible results in each case.
The combination of \apsq and electrostatic 3D TCAD  is therefore suitable for a range of different devices without the need for prior adaptation of the simulation set-up to the sensor design.

\begin{figure*}[bpt]
	\centering
	\begin{subfigure}[t]{0.5\textwidth}
		\centering
		\includegraphics[width=\columnwidth]{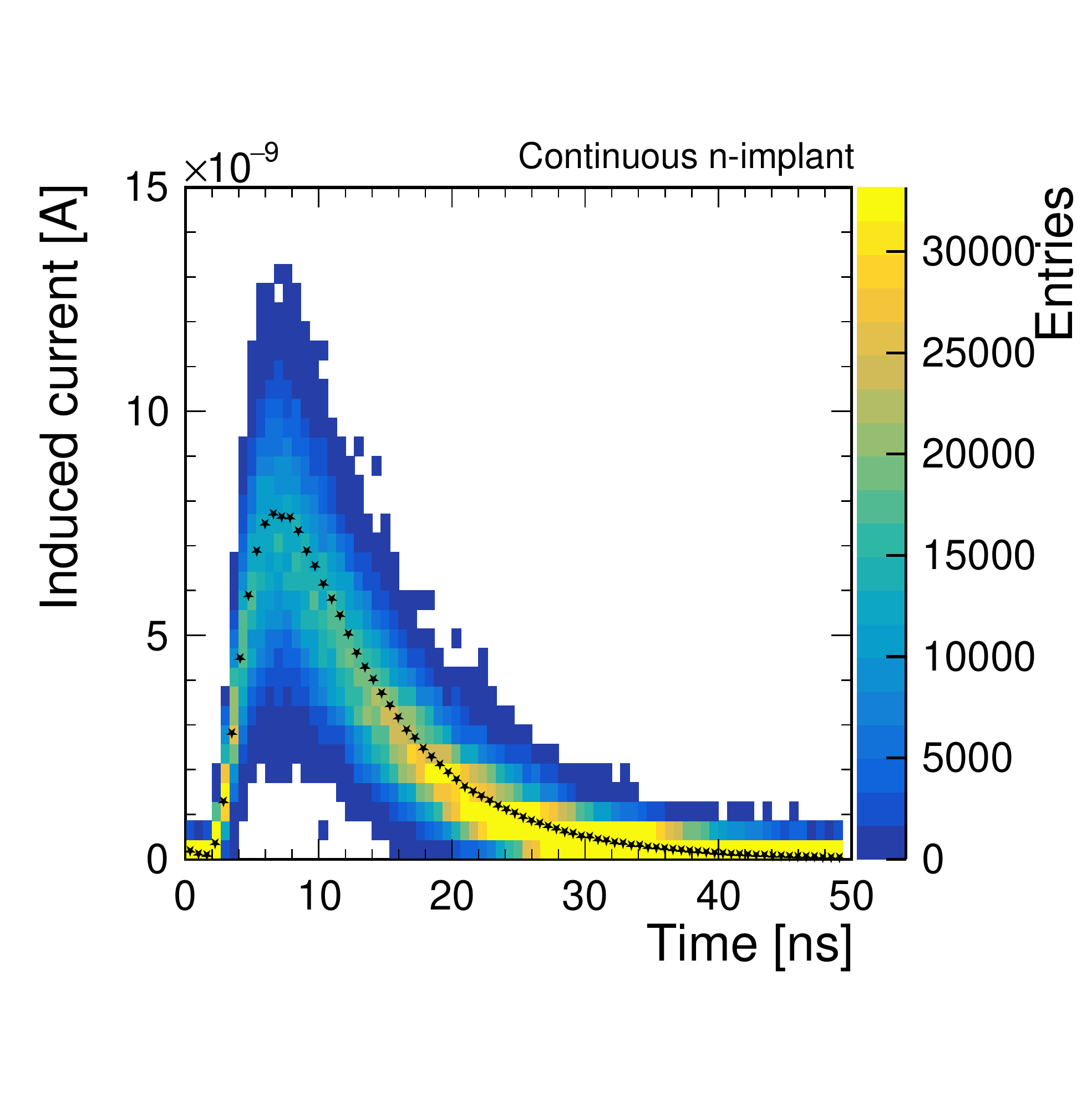}
		\caption{Transient sensor response}
		\label{fig:contN_densityPulse:diff}
	\end{subfigure}%
	\begin{subfigure}[t]{0.5\textwidth}
		\centering
		\includegraphics[width=\columnwidth]{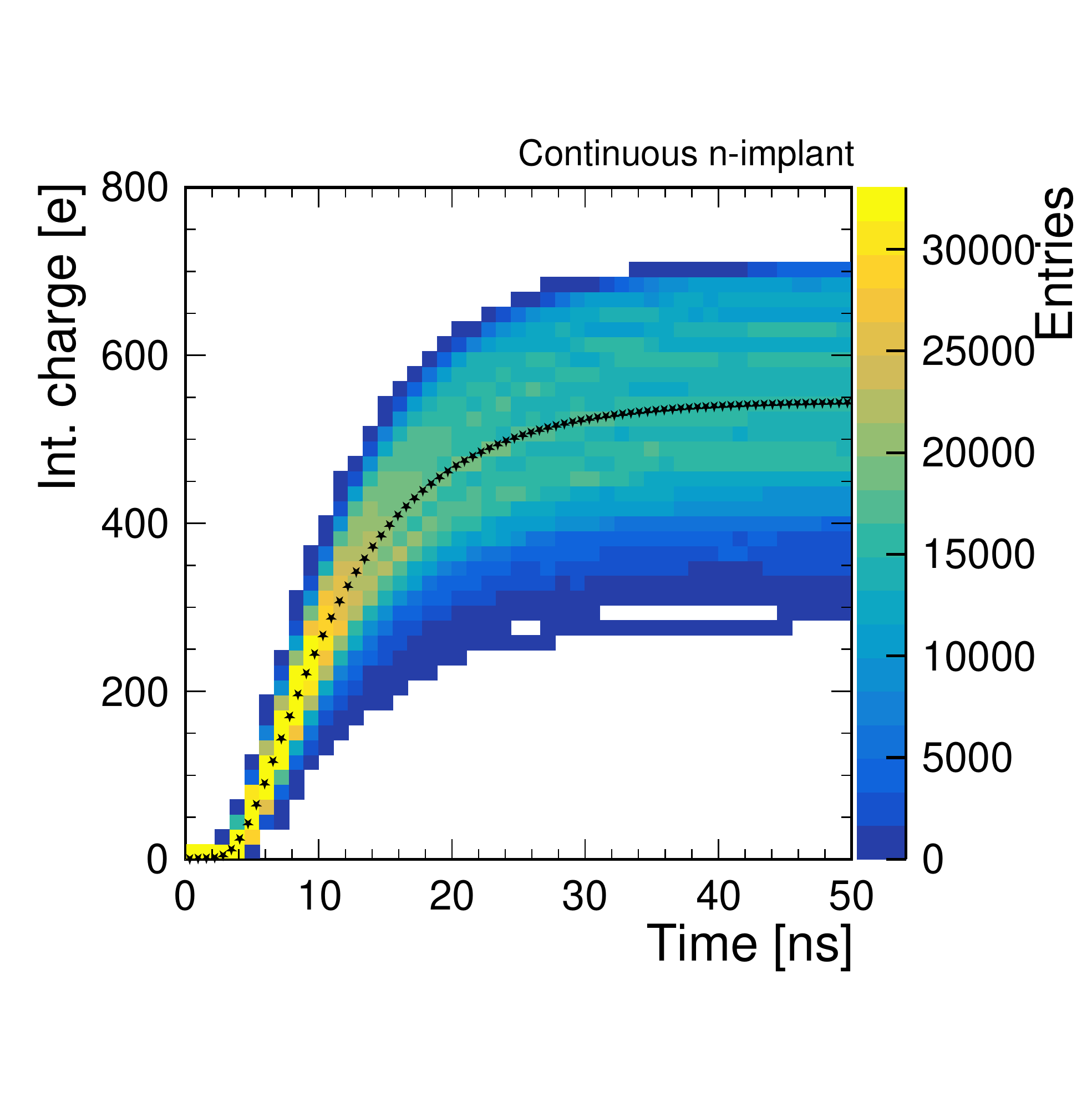}
		\caption{Integrated charge}
		\label{fig:contN_densityPulse:int}
	\end{subfigure}%
	\caption{(a) Transient pulse distribution and (b) integrated charge for the pixel design with continuous n-implant, obtained from a full Monte Carlo simulation of particles impinging at the pixel corner.
	The simulation includes secondary particles and energy deposition fluctuations. The black stars represent the average pulse height and integrated charge, respectively.}
	\label{fig:contN_densityPulse}
\end{figure*}

The high simulation rates in \apsq allow the simulation to be repeated while varying the charge injection positions systematically over the pixel cell.
In this way, the entire pixel cell can be probed, as shown in Fig.~\ref{fig:transientPulseModCornerAllSensor}, where the average induced current pulse can be seen for the pixel design with continuous n-implant.
For the sensor average, 1200 events are simulated to obtain a homogeneous charge injection across the pixel cell.
The peak of the average transient pulse for the full sensor is located below one nanosecond, in contrast to the peak at 7-8\,ns for charge injection at the pixel corner, which represents the worst case in terms of sensor time performance owing to the low lateral electric field in this region.

Statistical fluctuations of the charge deposition as well as the generation of secondary particles such as delta rays can now also be taken into account with reasonable statistics.
In \apsq this can be achieved by switching the charge deposition stage of the simulation to the \emph{DepositionGeant4} module, using the configuration presented in ~\cite{allpix-hrcmos}.

The transient pulse distribution for particles incident on the pixel corners is displayed in Fig.~\ref{fig:contN_densityPulse:diff} and their integrated induced current in Fig.~\ref{fig:contN_densityPulse:int}.
Here, the black stars indicate the average pulse height value and integrated charge in the respective time bin, respectively.
The pulse-by-pulse variations underline the importance of including statistical effects in the simulation setup.

%% file: time_resolution.tex
\subsection{Sensor Time Resolution Studies}

\begin{figure*}[tb]
	\centering
	\begin{subfigure}[t]{0.5\textwidth}
		\centering
		\includegraphics[width=\columnwidth]{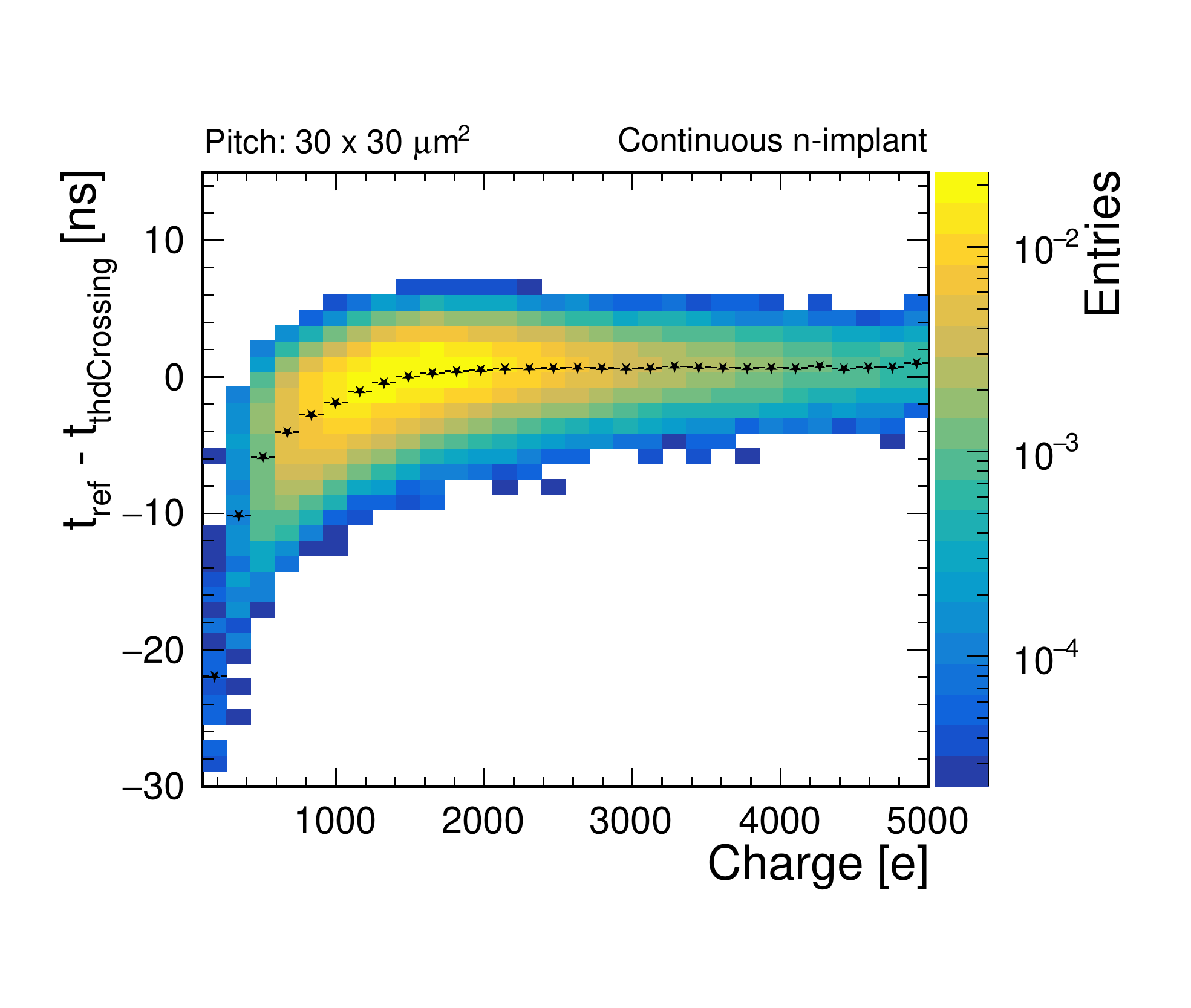}
		\caption{Time residuals before correction.}
		\label{fig:contScatterMean:Uncorrected}
	\end{subfigure}%
	\begin{subfigure}[t]{0.5\textwidth}
		\centering
		\includegraphics[width=\columnwidth]{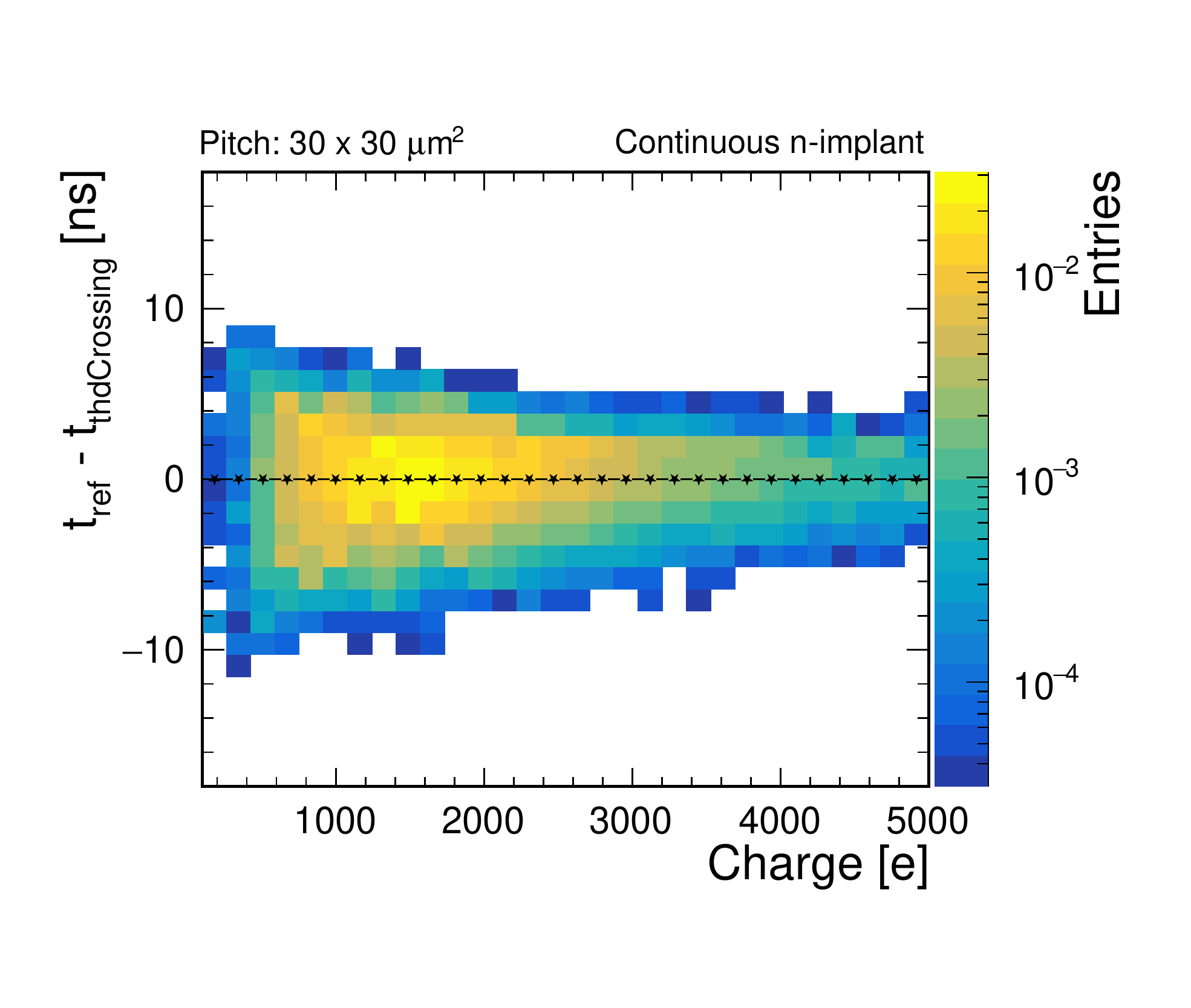}
		\caption{Time residuals after correction.}
		\label{fig:contScatterMean:Corrected}
	\end{subfigure}
	\caption{Time residuals as a function of the highest single pixel charge in a cluster before~(\subref{fig:contScatterMean:Uncorrected}) and after~(\subref{fig:contScatterMean:Corrected}) correction of the charge-dependent threshold crossing time. The black stars denote the mean of each charge bin.}
\end{figure*}

The transient Monte Carlo simulation validated in the previous section allows for an evaluation of the sensor time resolution already at the prototyping stage, which is crucial for sensor optimisation studies.
The inclusion of all relevant statistical elements of the signal formation in the sensor enable a realistic estimation of the sensor response time with different operation conditions and with different sensor designs.
Here, this approach is exemplified by evaluating the sensor time resolution for different pixel pitches and the sensor design with continuous n-implant.

Similar to the functioning of many detector front-ends, the time at which the current pulse crosses
the chosen detection threshold (\textit{threshold crossing time}) is used to compute the time residuals of the sensor.
The crossing time obtained from the simulation is smeared by a Gaussian distribution with a standard deviation of \SI{1}{ns} in order to account for the resolution of a time reference measurement such as a beam telescope or a trigger scintillator.

Fig.~\ref{fig:contScatterMean:Uncorrected} shows the time residual distribution as a function of the signal charge for a sensor with a pixel pitch of \SI{30x30}{\micro m}.
Small transient pulses cross the threshold later, which leads to a tail observable for low signal values.
This charge-dependence in threshold crossing time is corrected by subtracting the mean time offset for each charge bin separately, yielding the corrected distribution displayed in Fig.~\ref{fig:contScatterMean:Corrected}.
The observed behaviour matches very well the timewalk effect known from measurements.
The simulated time however only comprises the effects related to the signal formation in the sensor such as longer charge collection times in the pixel corners but not the additional contributions from e.g.\ threshold fluctuations in the front-end electronics.

\begin{figure}[tbp]
	\centering
	\includegraphics[width=\columnwidth]{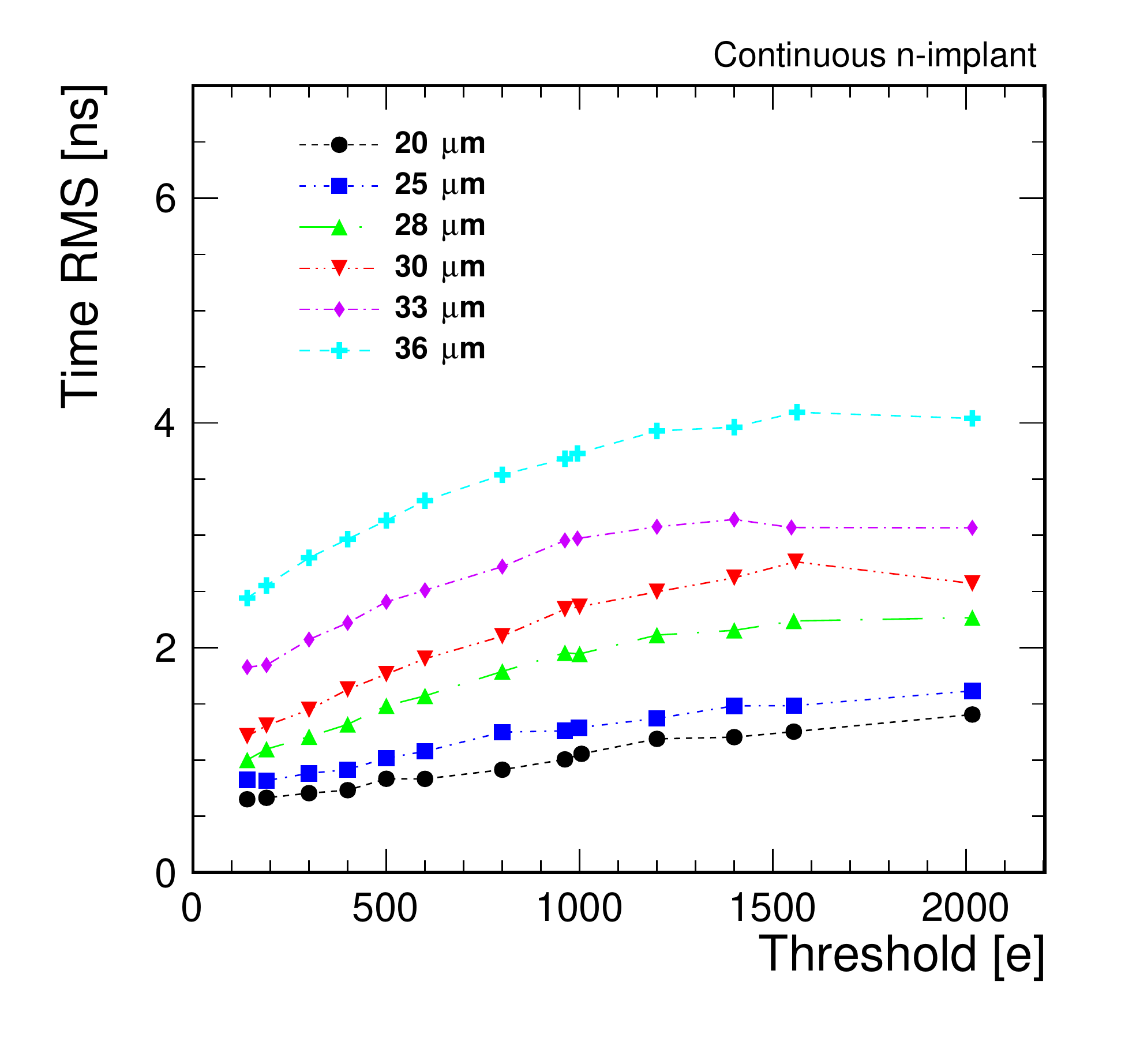}%
	\caption{Time resolution as a function of the detection threshold for a simulated sensor with different pixel pitches.}
	\label{fig:timeVsThresholdModFlavour}
\end{figure}

Using this corrected time residual distribution, the time resolution of the sensor can be extracted and the width of the distribution can be compared between different prototype designs.
Here, the width of the time residual distribution is quoted as the RMS of the central $3\sigma$.
Fig.~\ref{fig:timeVsThresholdModFlavour} shows the width of the distribution as a function of the applied detection threshold for different pixel pitches.
As expected, the time resolution improves for smaller pixel pitches due to a more homogeneous time response across the pixel cell.
With increasing detection threshold, a deterioration of the time resolution can be observed, owing to the flattening shape of the signal resulting in a stronger contribution to time jitter.  

%% file: reco_analysis.tex
In the following sections the transient Monte Carlo simulation using electrostatic TCAD fields and \apsq is compared to data recorded with the CLICTD prototype in test-beam measurements.
This section summarises the reconstruction and analysis techniques, while a detailed description of the setup and reconstruction is given elsewhere both for the test-beam measurements~\cite{clictdTestbeam} and for the simulations~\cite{allpix-hrcmos}.

\subsection{Test-beam Measurements}

The test-beam measurements presented were performed at the DESY~II Test~Beam Facility~\cite{Diener:2018qap} using a \SI{5.4}{GeV} electron beam.

\paragraph{Experimental Setup}

An EUDET-type telescope featuring six planes of MIMOSA\,26 monolithic active pixel sensors~\cite{Jansen:2016bkd} and an additional Timepix\,3 time-reference plane~\cite{Poikela_2014} were used for reference measurements of particle tracks.
The CLICTD sensor was placed as Device Under Test (DUT) between three MIMOSA\,26 planes in the upstream arm and three in the downstream arm of the telescope.
The Timepix3 plane was located downstream of the last MIMOSA\,26 plane.

All data presented in the subsequent section has been recorded with a CLICTD prototype featuring the continuous n-implant sensor design, a pixel pitch of \SI{37.5x30}{\micro m} and a total thickness of \SI{100}{\micro m}.

\paragraph{Reconstruction}

The offline reconstruction is performed with the Corryvreckan framework~\cite{corry_paper}.

First, adjacent pixels on each of the telescope planes and on the DUT are combined into clusters.
The cluster position is given by a centre-of-gravity algorithm using charge-weighting where applicable.
The cluster position on the DUT is corrected with the $\eta$-formalism~\cite{Akiba:2011vn} to take non-linear charge sharing into account.

Track candidates are required to have a cluster on each telescope plane, the  DUT is excluded from tracking.
The General Broken Lines formalism~\cite{Blobel:2006yi} is used to fit the tracks.
The resolution of the track impact position has been determined to be between $\SI{2.4}{\micro m}$ and $\SI{2.8}{\micro m}$ for the measurements featuring a particle beam perpendicular to the sensor surface~\cite{resolutionSimulator}.
For measurements with a rotated DUT, the resolution is $\SI{5.7}{\micro m}$ since the distance between the telescope planes and the DUT is enlarged to enable rotation of the sensor.

A reconstructed track is associated to a DUT cluster if the distance between the track incidence position on the DUT and the cluster position is less than 1.5 times the pixel pitch, and the track time falls within the active shutter of the DUT.

The hit detection efficiency is defined as the number of associated tracks divided by the total number of tracks, provided that the tracks pass trough the acceptance region of the DUT.
The acceptance region is defined as the physical pixel matrix excluding the outermost row/column as well as masked pixels and their direct neighbours.
The number of masked pixels on the DUT is below \SI{0.1}{\percent}.

\paragraph{Systematic Uncertainties}

The detection threshold applied to the sensor has a considerable impact on the cluster observables and has been calibrated in laboratory measurements~\cite{clictdTestbeam}.
The calibration yields a statistical uncertainty of $\pm \SI{0.02}{}$\,e$^-$/threshold DAC step  and a systematic uncertainty of~$^{+ 2.2}_{-3.0}\si{e}^-$/threshold DAC step.

The resolution of the reference telescope was found to have a systematic uncertainty of $\pm \SI{0.1}{\micro \meter}$ originating from the uncertainty on the intrinsic resolution of the MIMOSA26 telescope planes~\cite{Jansen:2016bkd}.

\subsection{Simulation}

A particle beam consisting of electrons with a momentum of 5.4\,GeV is simulated in order to replicate the test-beam conditions.
In total, approximately 2.1 million primary events for each diffusion configuration were simulated and have been reprocessed for every threshold setting, which took about 10\,h using 8 threads. 

Instead of reconstructed particle tracks, the Monte Carlo truth information is used as a reference.
The reference position on the DUT is calculated by linearly interpolating between the entry and exit point of the particle in the sensor.
The position as well as the simulated time measurements are smeared by a Gaussian distribution with a standard deviation corresponding to the respective track resolution of the corresponding data run.

The same reconstruction techniques used for the test-beam data are applied.

%% file: uncertainties.tex
\paragraph{Systematic Uncertainties}
\label{sec:uncertainties}

The spatial and time granularity of both the 3D TCAD and the Monte Carlo simulation needs to be fine enough to resolve the relevant physical features and processes.
The time granularity of the simulated charge carrier transport in \apsq is set to \SI{7}{ps}.
The value has been varied from $\SI{1}{ps}$ to $\SI{20}{ps}$ and the impact on the final observables was found to be negligible.
The 3D TCAD simulation applies an adaptive spatial and time granularity, using a fine-grained mesh for small feature sizes.
It has been verified that a further decrease in mesh size and time granularity does not alter the results.
The spatial granularity of the 3D TCAD maps imported into \apsq is fixed to $\SI{0.2 x 0.2 x 0.2 }{\micro m}$.
A finer meshing has no significant effect on the final observables.

The charge carrier transport in \apsq is sped up by transporting charge carriers as a group.
Here, the number of charge carriers per group is set to 10.
The value was varied between 1 and 15 and no significant impact was observed.
The time interval in which the charge carriers are transported is fixed to \SI{50}{ns}.
It has been verified that continuing the simulation beyond the \SI{50}{ns} does not significantly alter the results for the pixel flavour with continuous n-implant confirming that the majority of liberated charge carriers have either recombined or reached the collection electrode.  
Smaller values lead to a decrease in the integrated induced signal, as expected.

The sensor properties were found to be highly sensitive to the doping profiles used to model the sensor.
The Spreading Resistance Profiling (SRP) technique was used to measure the resistivity of a \SI{18}{\micro m} thick epitaxial layer, that was scaled to \SI{30}{\micro m} to match the investigated sensor design~\cite{thesis-jacobus}.
The profile is characterised by a transition region between the high-resistivity epitaxial layer and the low-resistivity substrate, that arises from the diffusion of dopants out of the substrate into the epitaxial layer.
The slope of the transition region was varied by a factor of three to account for uncertainties in the SRP measurement and the scaling of the profile to \SI{30}{\micro m}.
For the collection electrode and the p-well implant, the vertical extent of the doping profiles are taken from process simulations performed by the foundry.
The lateral diffusion of these profiles is not known and is therefore modelled by a Gaussian function. 
The width of the Gaussian is varied by a factor of three to account for the uncertainty on the diffusion.  

The total uncertainty is determined by varying the lateral diffusion at the p-well and at the collection electrode as well as the vertical diffusion of the substrate individually and repeating the simulation.
The different sources of uncertainty are assumed to be uncorrelated and for each observable, the residuals between the simulations with nominal values and the ones with varied doping profiles are determined and then summed quadratically.

%% file: data_comparison.tex
In the following, different observables in the test-beam measurements and the simulation are compared at nominal operation conditions with a charge threshold of approximately \SI{150}{e}, as well as in a threshold scan applying charge thresholds between \SI{100}{e} and \SI{1800}{e}.

\begin{table*}[tbp]
	\centering
	\begin{tabular}{lrr}
		\toprule
		\textbf{Parameter} & \textbf{Data} & \textbf{Simulation} \\
		\midrule
		Total cluster size & $1.94 \pm 0.01$ &  $2.00^{+ 0.05}_{- 0.12}$  \\
		Column cluster size & $1.36 \pm 0.01$ &  $1.37^{+ 0.03}_{- 0.04}$   \\
		Row cluster size & $1.44 \pm 0.01$  & $1.49^{+ 0.06}_{- 0.07}$  \\
		Spatial resolution (row) $\sigma^{(s)}_\textrm{row}$& $4.4\pm0.2$\,\si{\micro \meter}  & $4.3^{+ 0.43}_{- 0.04}$\,\si{\micro \meter} \\
		\bottomrule
	\end{tabular}
	\caption{Mean cluster size and spatial resolution in row direction for data and simulation at a detection threshold of 150\,e}
	\label{tab:tb_results}
\end{table*}

\subsection{Cluster Size}
\label{sec:data_comparison:size}

The cluster size is strongly influenced by charge sharing between neighbouring pixel cells.
It is therefore well-suited to study the electric field especially around the pixel edges and in the corners.

\begin{figure}[tbp]
	\centering
	\includegraphics[width=\columnwidth]{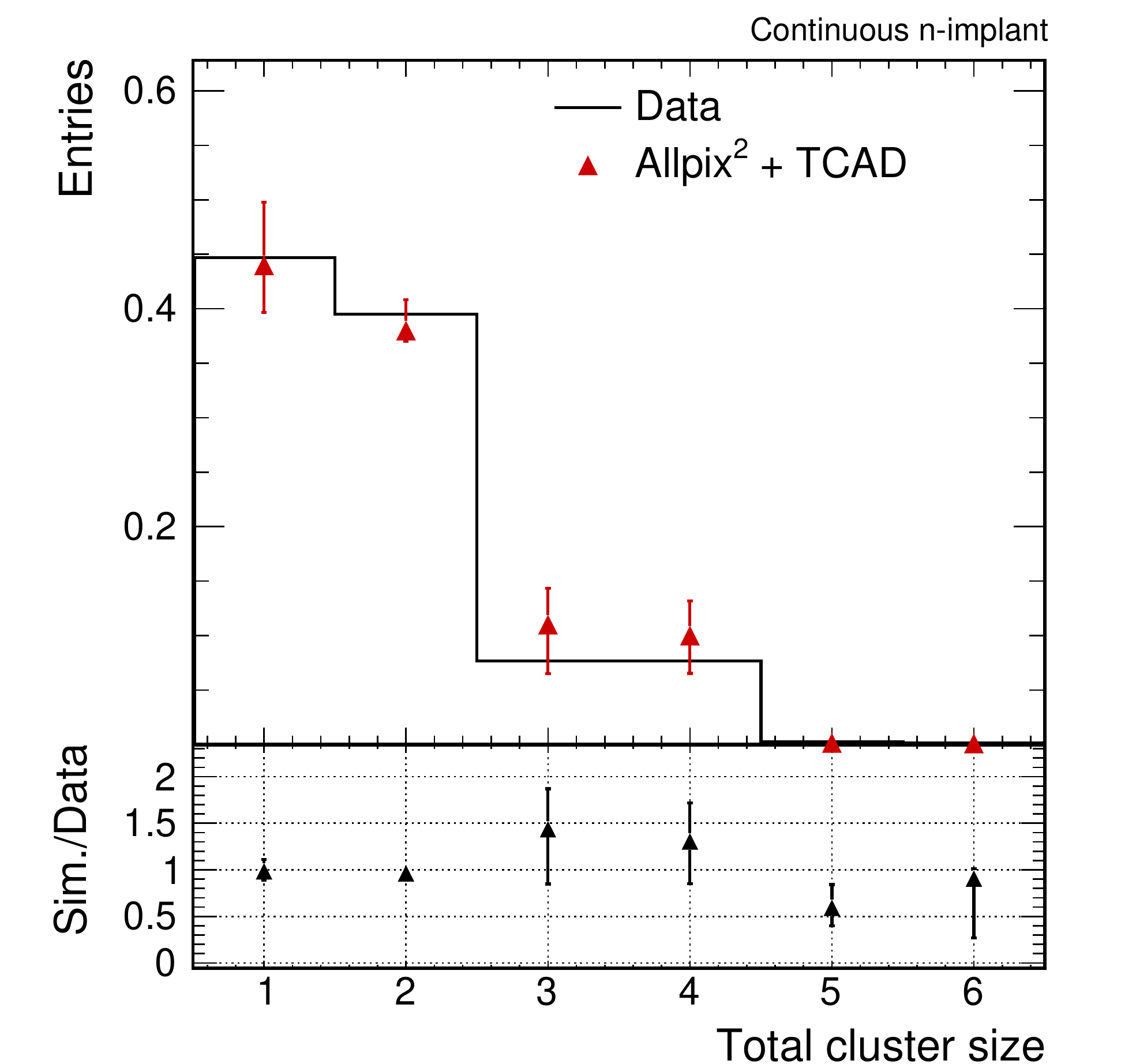}%
	\caption{Cluster size distributions of the total cluster size at a threshold of 150\,e.}
	\label{fig:clusterSize}
\end{figure}

\begin{figure}[tbp]
	\centering
	\includegraphics[width=\columnwidth]{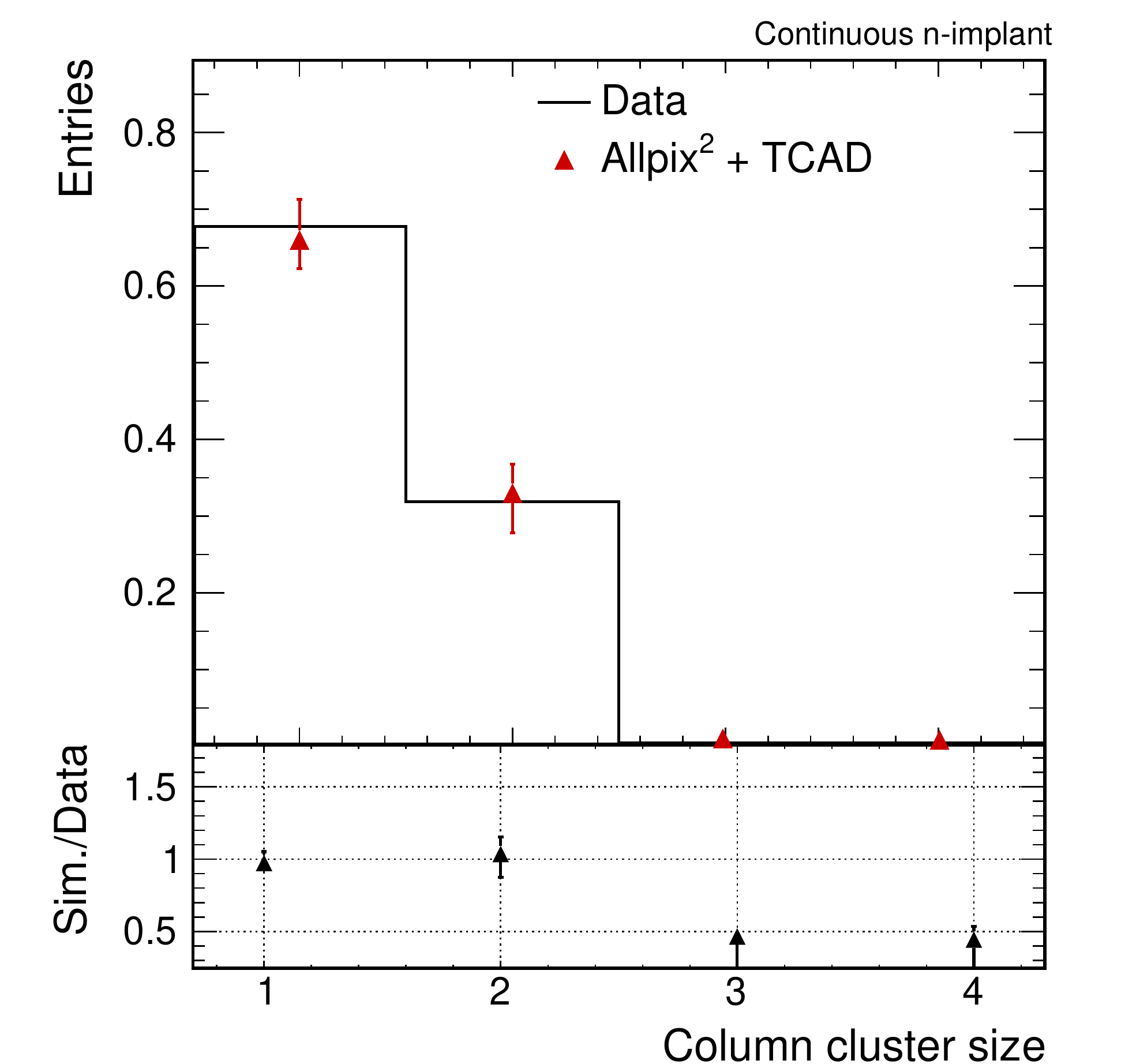}%
	\caption{Cluster size distribution in the column direction at a threshold of 150\,e.}
	\label{fig:clusterSizeX}

\end{figure}
\begin{figure}[tbp]
	\centering
	\includegraphics[width=\columnwidth]{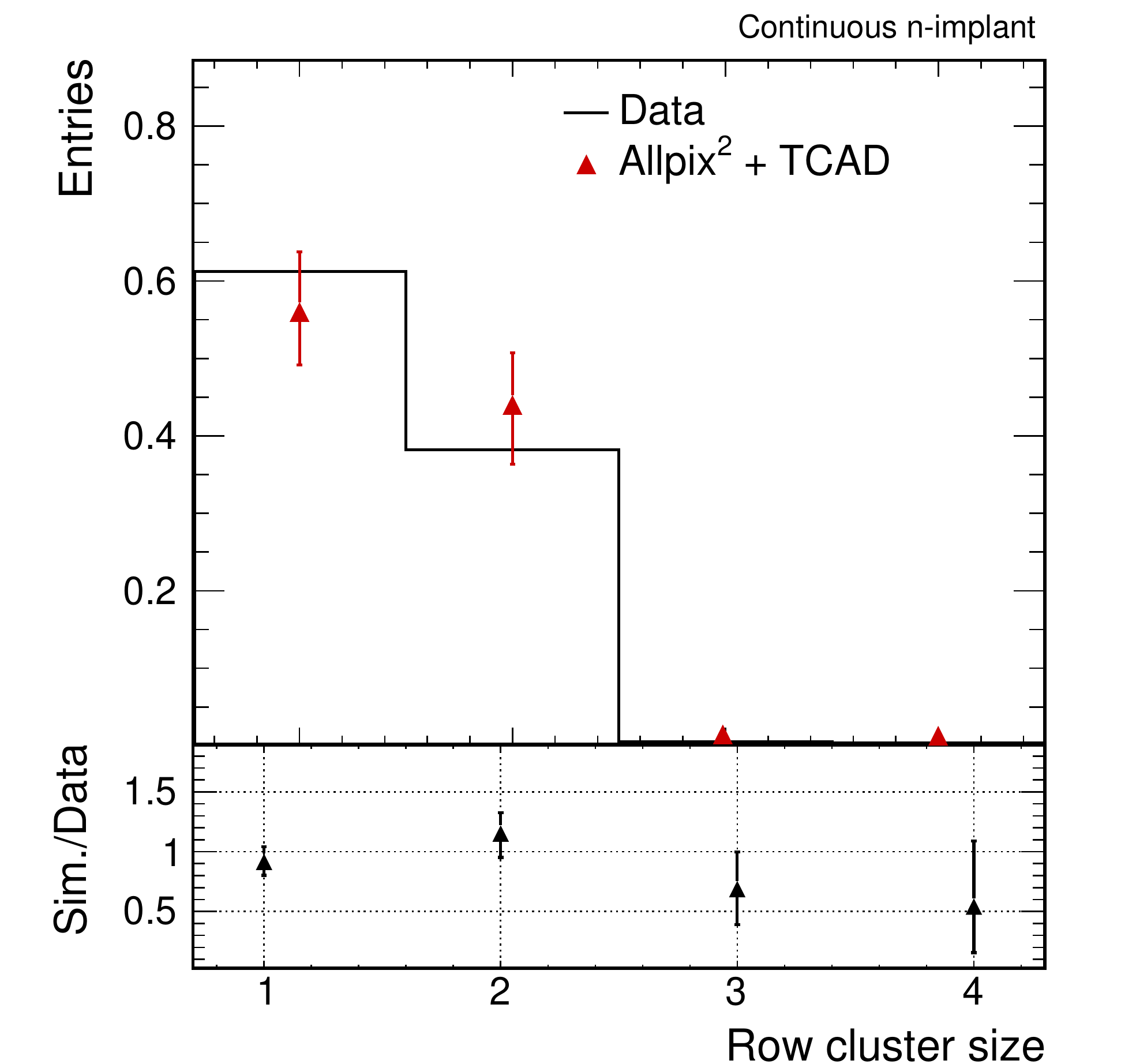}%
	\caption{Cluster size distribution in the row direction at a threshold of 150\,e.}
	\label{fig:clusterSizeY}
\end{figure}

\paragraph{Nominal conditions}
Fig.~\ref{fig:clusterSize} compares the total cluster size distribution observed in data and simulation at the nominal operation threshold.
The projections onto the column and row dimensions are shown in Fig.~\ref{fig:clusterSizeX} and Fig.~\ref{fig:clusterSizeY}, respectively.
The indicated uncertainties correspond to the contributions from the lateral diffusion of the doping profiles.

The overall shape of the size distribution is well captured both for the total cluster size as well as the projections, and the differences between data and simulation are within the systematic uncertainties that dominate over the statistical ones.
It can be observed that the uncertainties on cluster size in the row direction obtained from simulation are larger than in the column direction.
This is a direct effect of the shorter pixel pitch in that direction and the resulting stronger effect of the lateral electric field components in the edge region on the charge sharing behaviour.

The results are summarised in Table~\ref{tab:tb_results}.
The statistical uncertainty is of the order of $10^{-4}$ for both data and simulation.

\begin{figure*}[tbp]
	\centering
	\begin{subfigure}[t]{0.5\textwidth}
		\centering
		\includegraphics[width=\columnwidth]{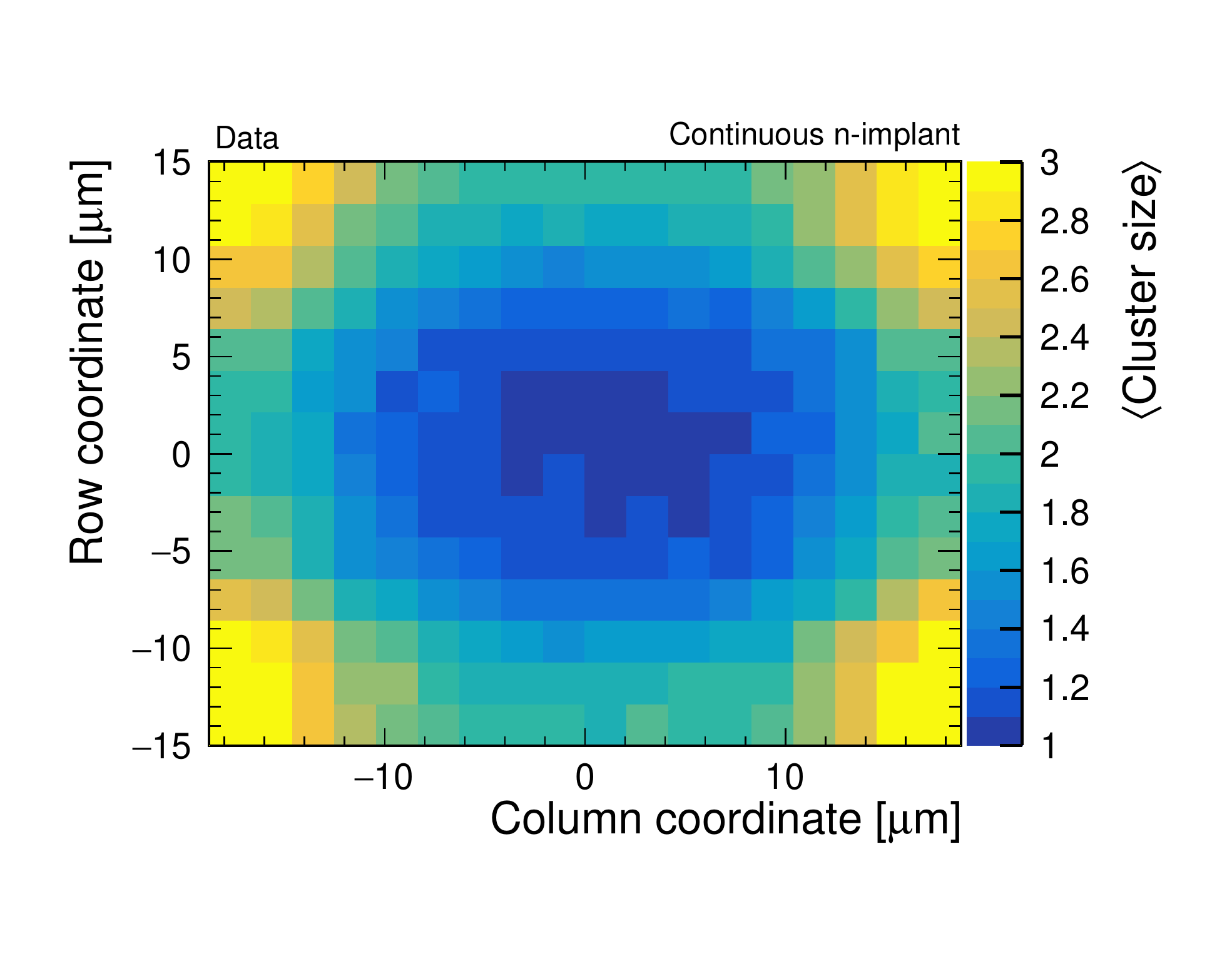}%
		\caption{Intra-pixel cluster size -- Data}
		\label{fig:sizeMap:Data}
	\end{subfigure}%
	\begin{subfigure}[t]{0.5\textwidth}
		\centering
		\includegraphics[width=\columnwidth]{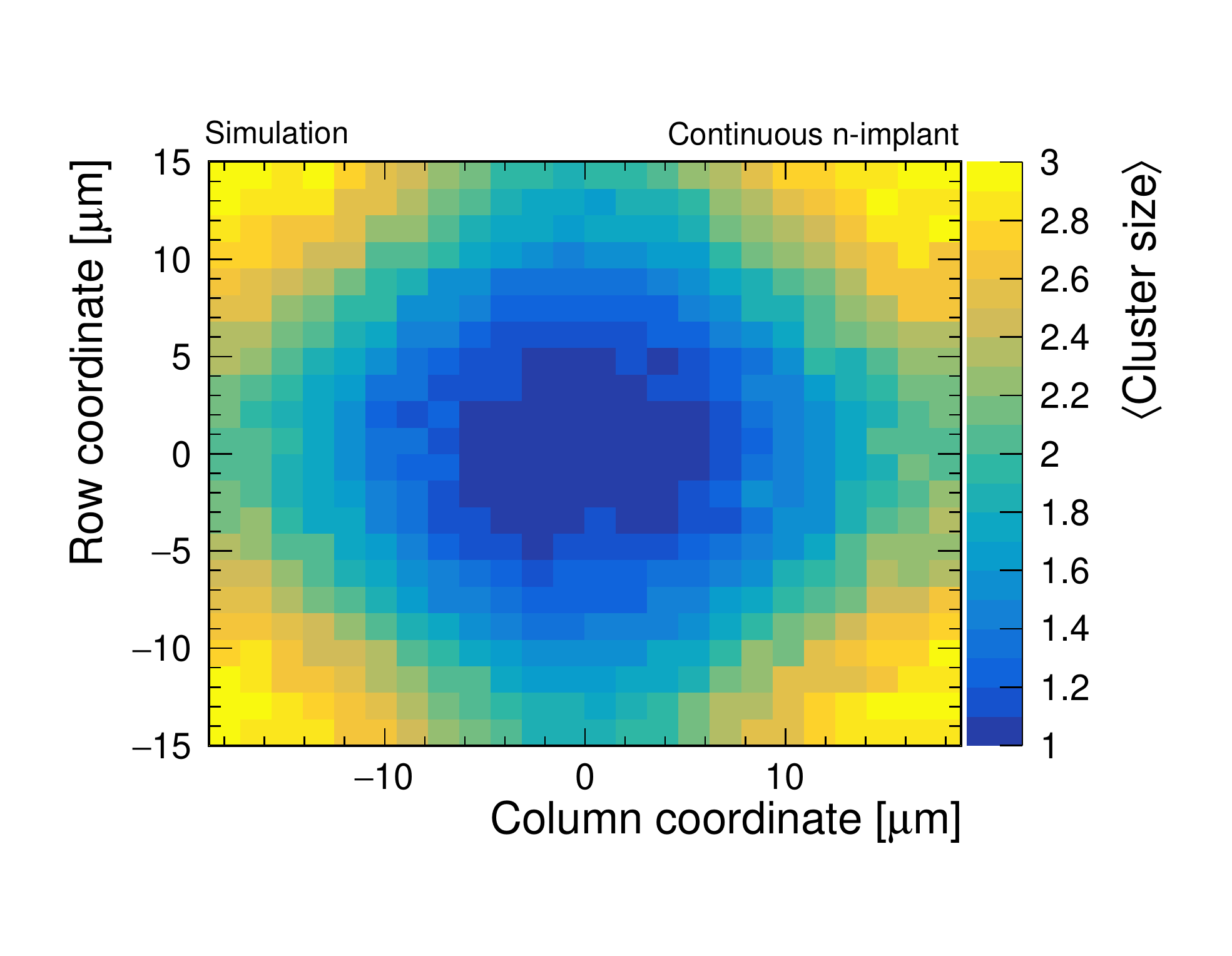}%
		\caption{Intra-pixel cluster size -- Simulation}
		\label{fig:sizeMap:Sim}
	\end{subfigure}%
	\caption{Intra-pixel distribution of the total cluster size as a function of the particle incidence position within the pixel cell for data~(\subref{fig:sizeMap:Data}) and simulation~(\subref{fig:sizeMap:Sim}), both at a threshold of \SI{150}{e}. Shown is a single pixel cell of the CLICTD prototype with a pitch of \SI{37.5x30}{\micro m}.}
	\label{fig:sizeMap}
\end{figure*}

With the availability of both high-statistics Monte Carlo simulations and a high-resolution beam telescope in the measurement campaign, a comparison of the cluster size at the sub-pixel level can be performed.
The cluster size as a function of the incident position within a single CLICTD pixel cell is displayed in Fig.~\ref{fig:sizeMap:Data} for data and in Fig.~\ref{fig:sizeMap:Sim} for simulation with nominal diffusion.
This intra-pixel representation provides a finely resolved perspective on the cluster size in different regions of the pixel and therefore allows the origin of possible remaining differences between data and simulation to be identified.
Here, the binning of the data plot is driven both by the resolution of the reference particle tracks and by the limited statistics of the available measurements.

The overall agreement between the distributions is very good.
In comparison with data, the size of the corner regions with three-pixel charge sharing is slightly overestimated in simulation.
This is in accordance with the slightly higher fraction of three-pixel clusters observed for simulation in Fig.~\ref{fig:clusterSize} and indicates a possible overestimation of charge sharing in these border regions which likely results from differences in the electric field.

\paragraph{Threshold scan}

\begin{figure}[tbp]
	\centering
	\includegraphics[width=\columnwidth]{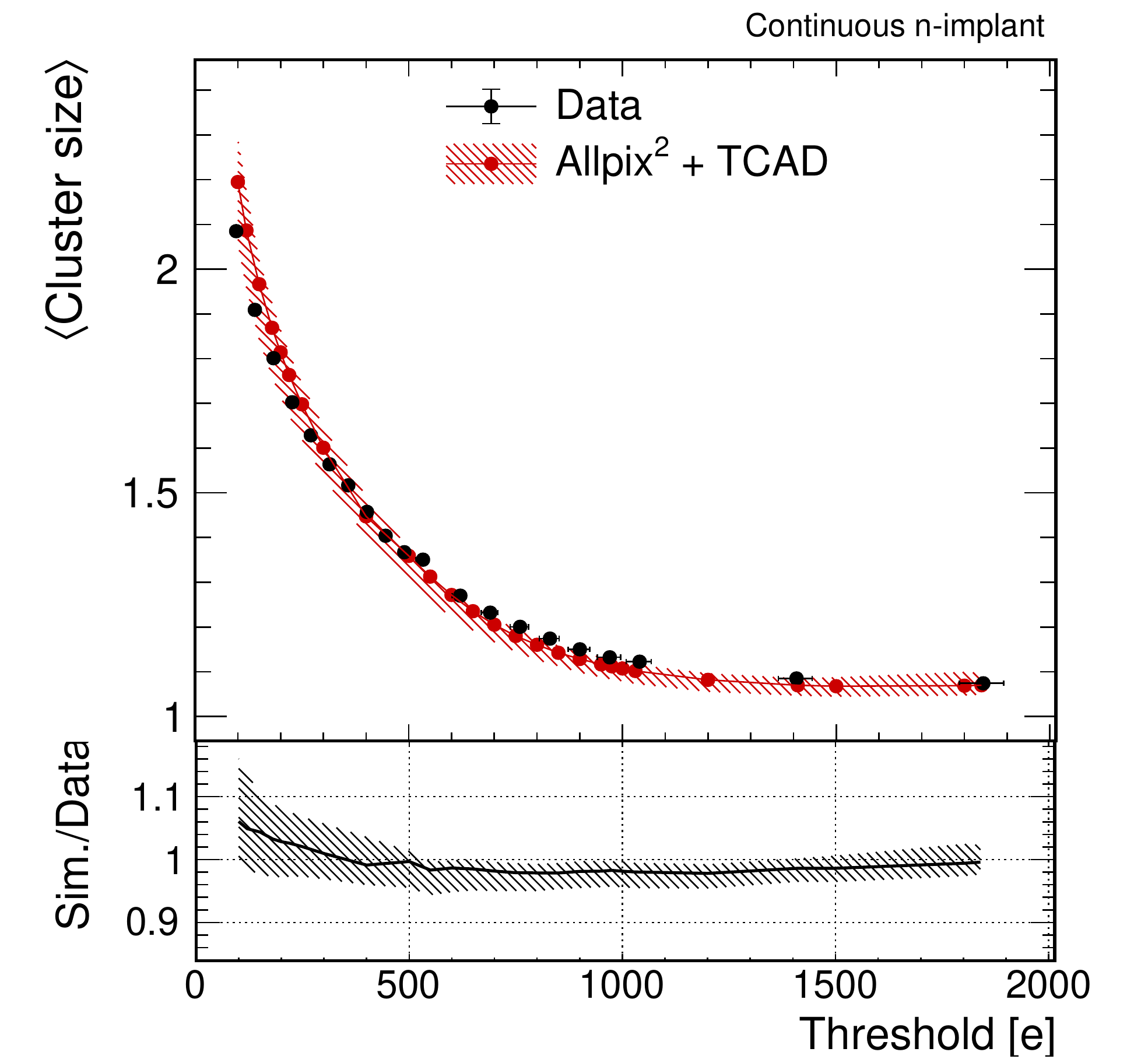}%
	\caption{Cluster size as a function of the detection threshold for data and simulation.}
	\label{fig:sizeThrdScan}
\end{figure}

\begin{figure}[tbp]
	\centering
	\includegraphics[width=\columnwidth]{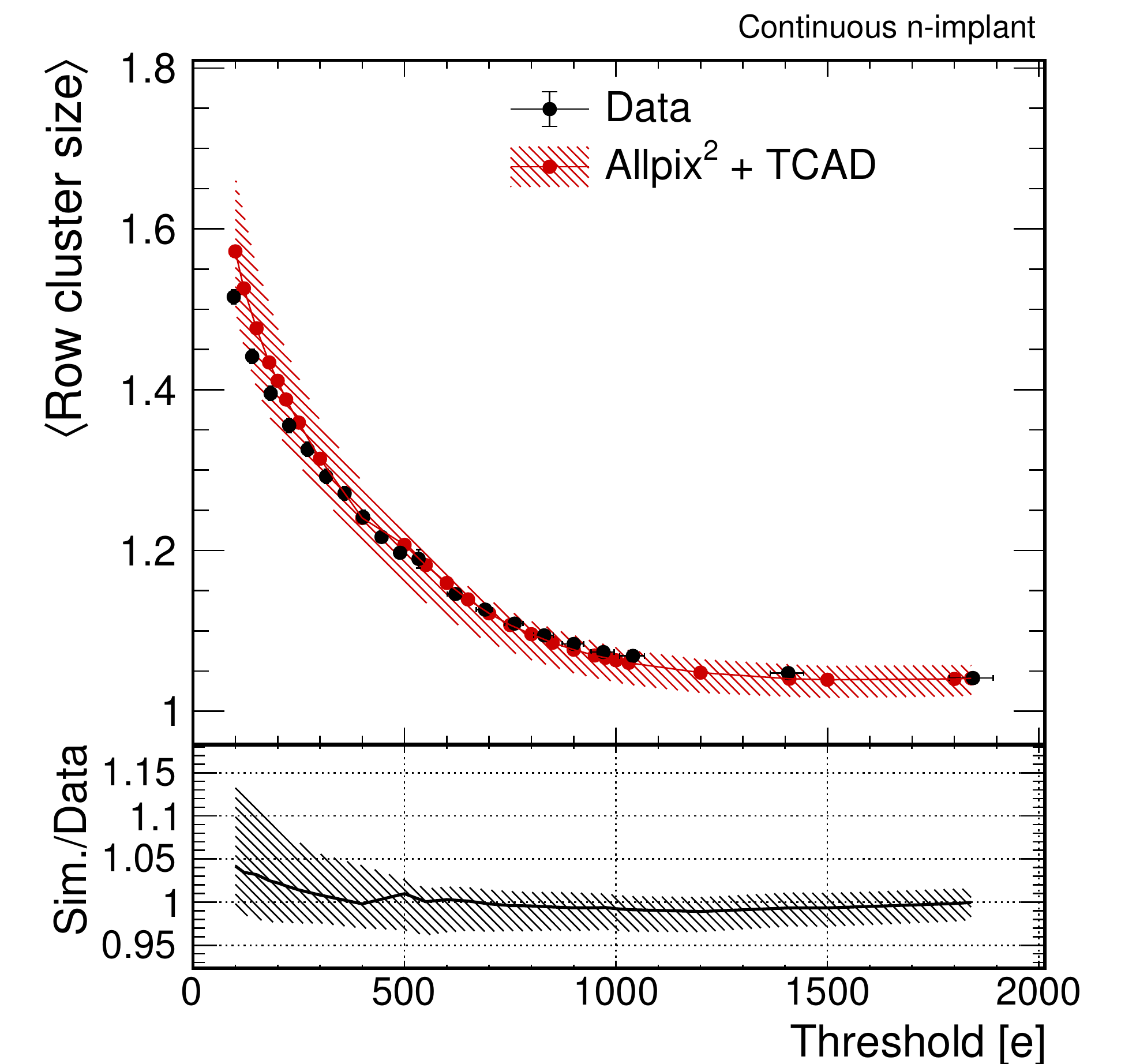}%
	\caption{Cluster size in row direction as a function of the detection threshold for data and simulation.}
	\label{fig:sizeYThrdScan}
\end{figure}

With increasing threshold the cluster size decreases with more and more pixels falling below the detection threshold.
Fig.~\ref{fig:sizeThrdScan} demonstrates this effect both for data and simulation, and it can be observed that the agreement between the two curves is well within the uncertainty over the full threshold range.
The same holds true for the projected cluster sizes as a function of the detection threshold, shown e.g. for the row direction in Fig.~\ref{fig:sizeYThrdScan}.
The maximum deviation is of the order of \SI{5}{\percent} for very low detection thresholds and covered by the systematic uncertainties.
This agreement over a wide range indicates that both the electric field and the charge propagation model replicates the physical situation in the sensor prototype sufficiently well.

\paragraph{Rotation scan}

\begin{figure}[tbp]
	\centering
	\includegraphics[width=\columnwidth]{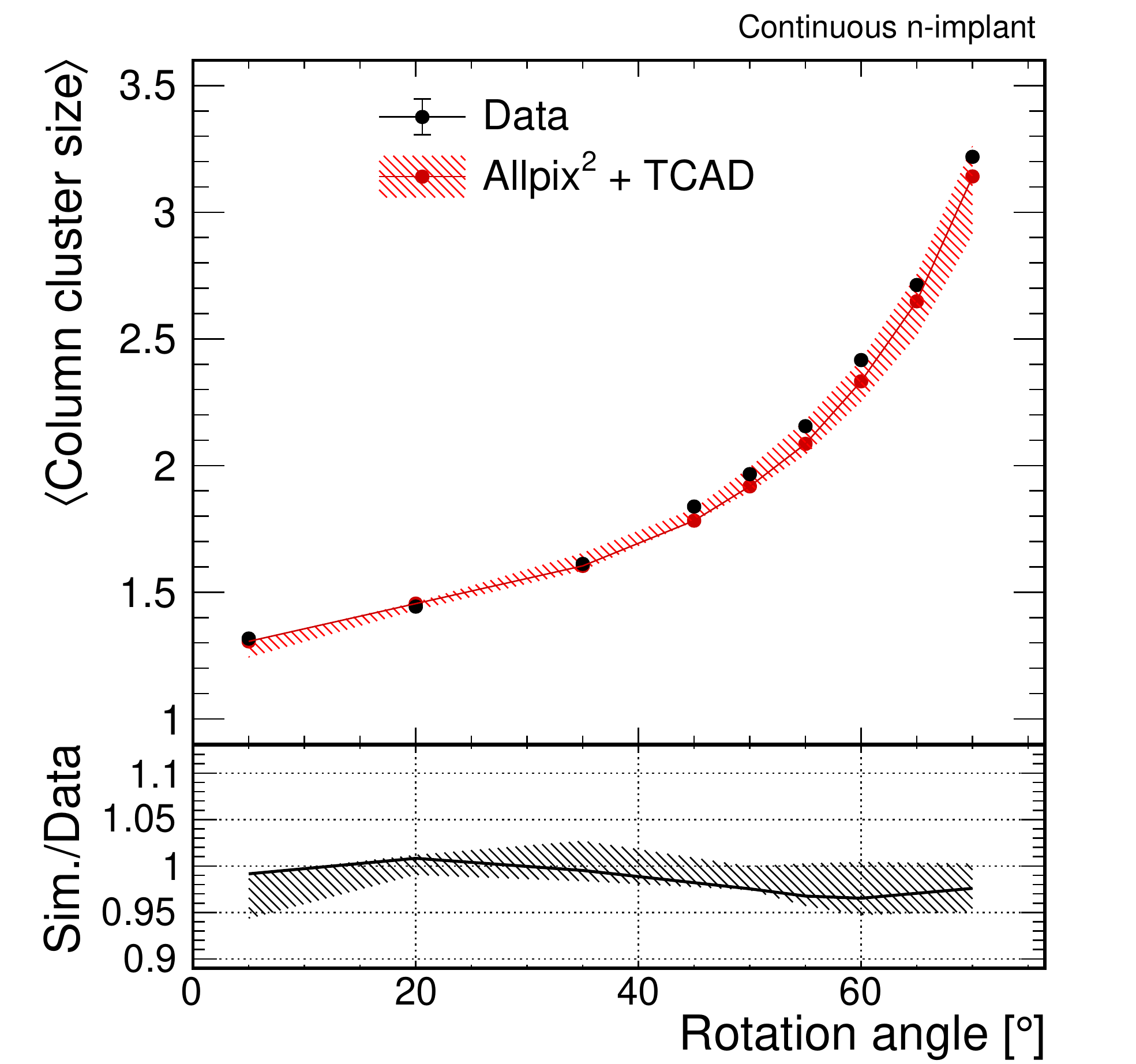}%
	\caption{Mean column cluster size as a function of rotation angle for data and simulation.}
	\label{fig:clsSizeX_vs_rotation}
\end{figure}

The agreement in charge sharing can be further probed by performing incident angle studies, in which the sensor is rotated relative to the particle beam and therefore the total path length traversed in silicon by a charged particle is changed.
With increasing rotation angles, the cluster size increases and charge carriers are created in different depths of the sensor in the pixels along the cluster.
Owing to the different electric field strength in the sensor regions as well as the different dominant effects of charge collection such as diffusion or drift, the size of the resulting clusters is very sensitive to the correct modelling of the signal formation.

For these measurements, the sensor was rotated around its vertical axis along a pixel column, and correspondingly an increase of the column cluster size is expected.
This is demonstrated in Fig.~\ref{fig:clsSizeX_vs_rotation}, where the mean column cluster size is shown as a function of rotation angle.

While the agreement is excellent for low rotation angles, a slightly larger deviation can be observed towards larger rotation angles that is still covered by the systematic uncertainty.
Here, additional signal contributions from the upper part of the substrate and the transition region between substrate and epitaxial layer have a significant impact on the cluster size, and the increasing difference can be attributed to the simplified modelling of this transition region.
The systematic deviation towards larger cluster sizes in data indicates that a lower diffusion in the simulation better describes the sensor under these operating conditions.

\subsection{Efficiency}

\begin{figure}[tb]
	\centering
	\includegraphics[width=\columnwidth]{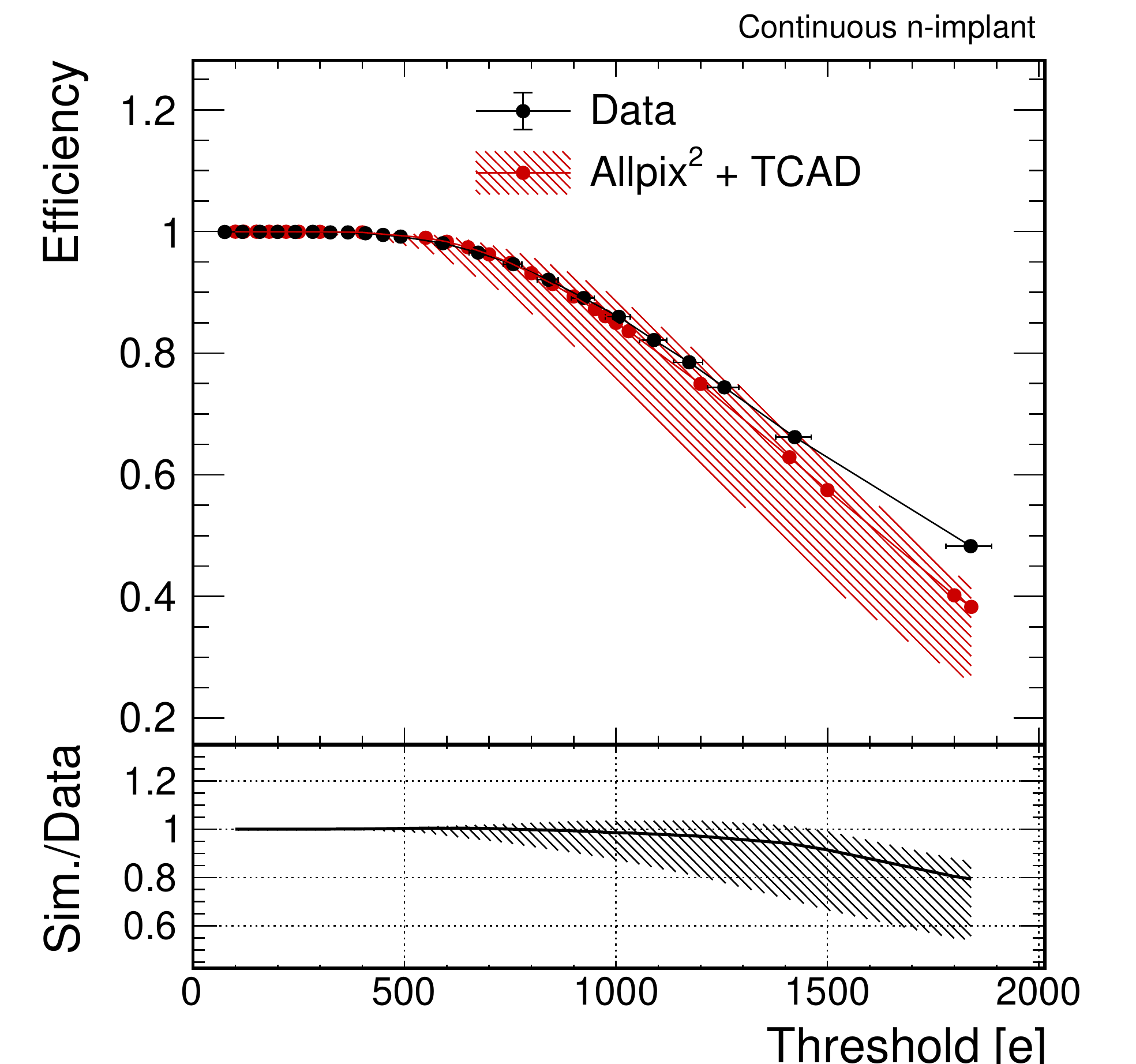}
	\caption{Detection efficiency as a function of the applied charge threshold for data and simulation.}
	\label{fig:overlayEfficiency}
\end{figure}

\begin{figure*}[tb]
	\centering
	\begin{subfigure}[t]{0.5\textwidth}
		\centering
		\includegraphics[width=\columnwidth]{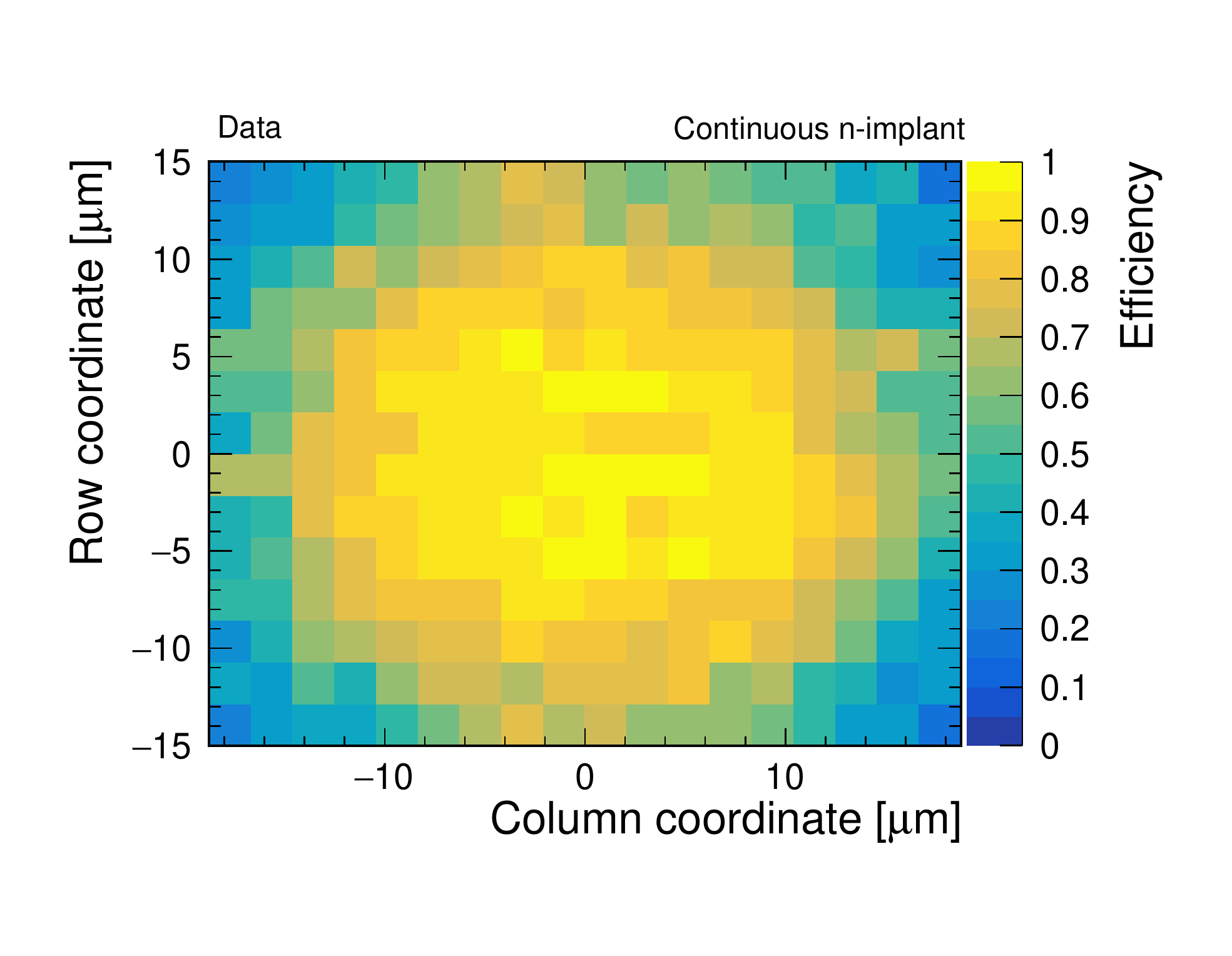}
		\caption{Intra-pixel efficiency -- Data}
		\label{fig:inPxlEfficiencyThrd1000:Data}
	\end{subfigure}%
	\begin{subfigure}[t]{0.5\textwidth}
		\centering
		\includegraphics[width=\columnwidth]{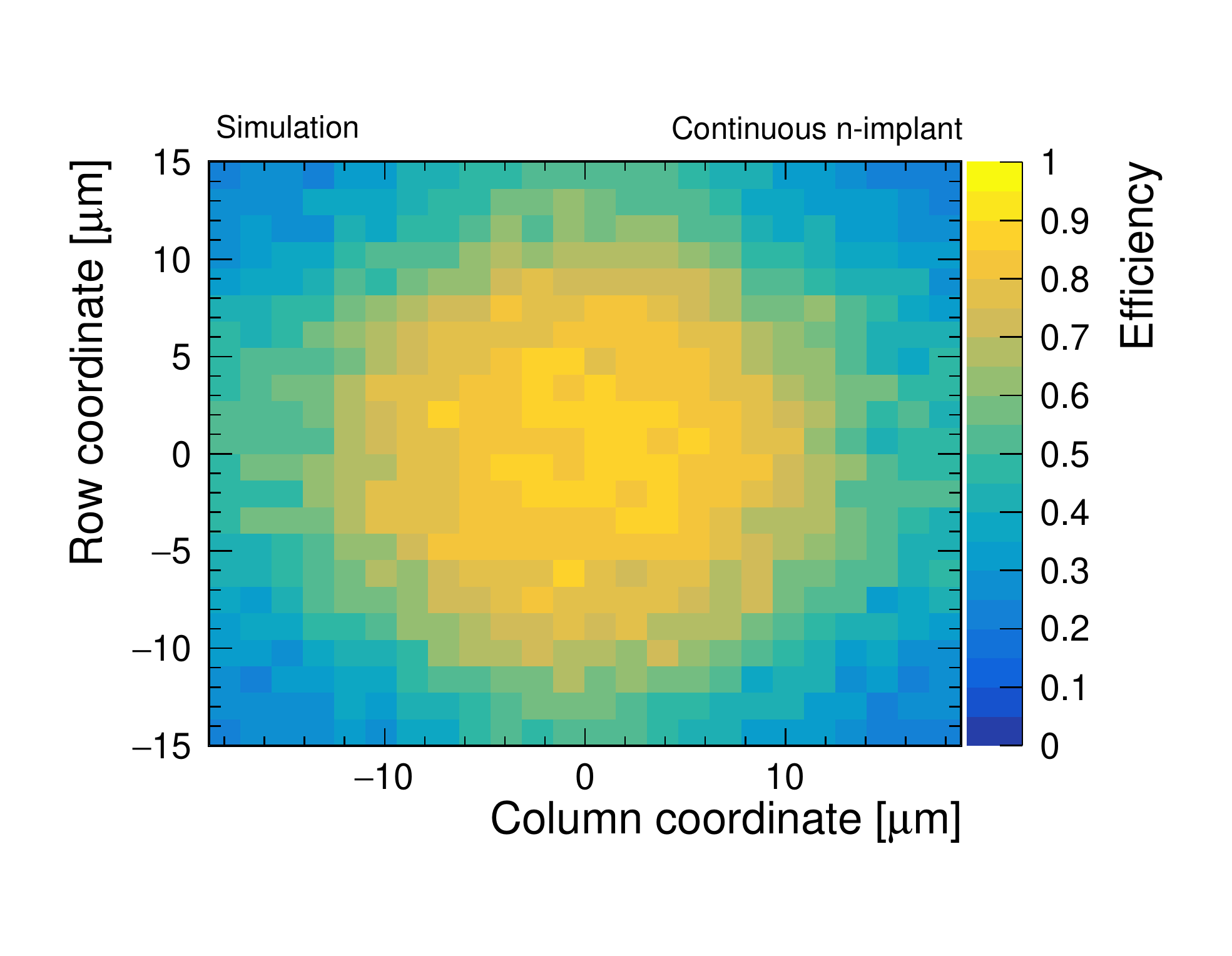}
		\caption{Intra-pixel efficiency -- Simulation}
		\label{fig:inPxlEfficiencyThrd1000:Sim}
	\end{subfigure}%
	\caption{Intra-pixel distribution of the detection efficiency as a function of the particle incidence position for data~(\subref{fig:inPxlEfficiencyThrd1000:Data}) and simulation~(\subref{fig:inPxlEfficiencyThrd1000:Sim}) at a detection threshold of approximately \SI{1850}{e}. Shown is a single pixel cell of the CLICTD prototype with a pitch of \SI{37.5x30}{\micro m}.}
	\label{fig:inPxlEfficiencyThrd1000}
\end{figure*}

A comparison of the efficiency as a function of the detection threshold in data and simulation is shown in Fig.~\ref{fig:overlayEfficiency}.

The efficiency is adequately reproduced for low detection thresholds, while data indicates a larger detection efficiency than predicted by simulation for very high thresholds.
From the distribution of the efficiency throughout the pixel cell at a threshold of approximately \SI{1850}{e}, shown in Fig.~\ref{fig:inPxlEfficiencyThrd1000:Data} and Fig.~\ref{fig:inPxlEfficiencyThrd1000:Sim} for data and simulation, respectively, no specific region can be identified that is responsible for the reduced efficiency observed in simulation.
However, it can be noticed that the efficiency drops more rapidly from the centre of the pixel cell towards the edges than observed in the data - an indication that either diffusion effects are overestimated or lateral electric fields are underestimated.
This is in agreement with the findings in Section~\ref{sec:data_comparison:size}.

\subsection{Spatial Resolution}

The spatial resolution achieved is calculated from the width of the residual distribution between reconstructed cluster position and the track position or the Monte Carlo particle position, for data and simulation, respectively.
For these studies, the width is defined as the reduced RMS of the central $3 \sigma$ (\SI{99.7}{\percent}) of the distribution.
To obtain the intrinsic sensor resolution, the track resolution is subtracted quadratically from the width.

\begin{figure}[tbp]
	\centering
	\includegraphics[width=\columnwidth]{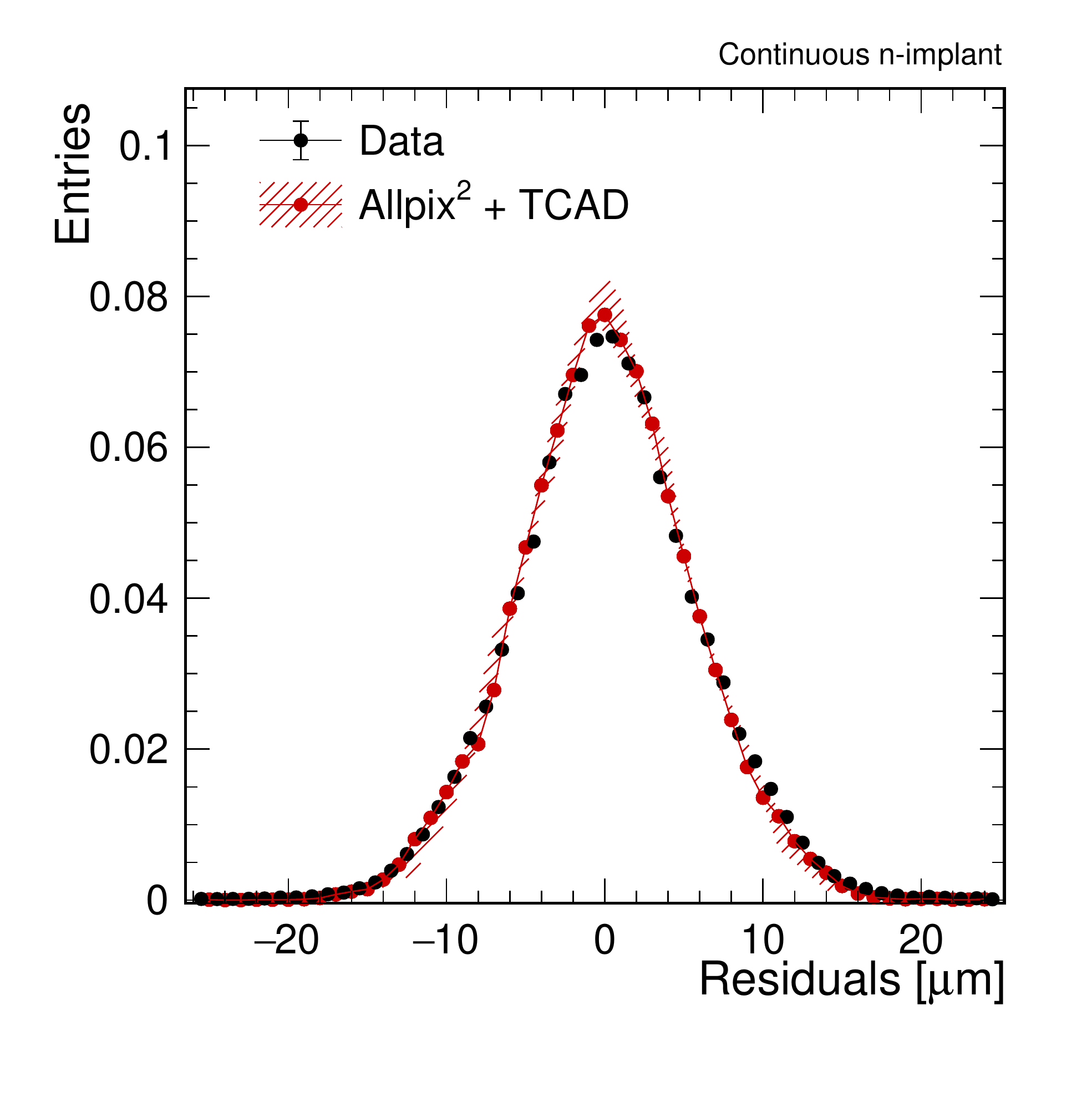}%
	\caption{Distribution of the spatial residuals between reconstructed DUT particle position and reference track for data and simulation at a threshold of \SI{150}{e}.}
	\label{fig:overlayResolutionY}
\end{figure}

\paragraph{Nominal threshold}
The residuals between the reconstructed cluster position and the particle incident position on the sensor are shown in Fig.~\ref{fig:overlayResolutionY} for row direction.
The width evaluates to \SI{5.3}{\micro m} for data and \SI{5.2}{\micro m} for simulation.
The telescope track resolution of $2.8 \pm 0.1$\,\SI{}{\micro m} is quadratically subtracted from the RMS, yielding a spatial resolution of $4.4 \pm 0.2$ \SI{}{\micro m} in data and $4.3^{+ 0.43}_{- 0.04}$ \SI{}{\micro m} in simulation.

\begin{figure}[tbp]
	\centering
		\includegraphics[width=\columnwidth]{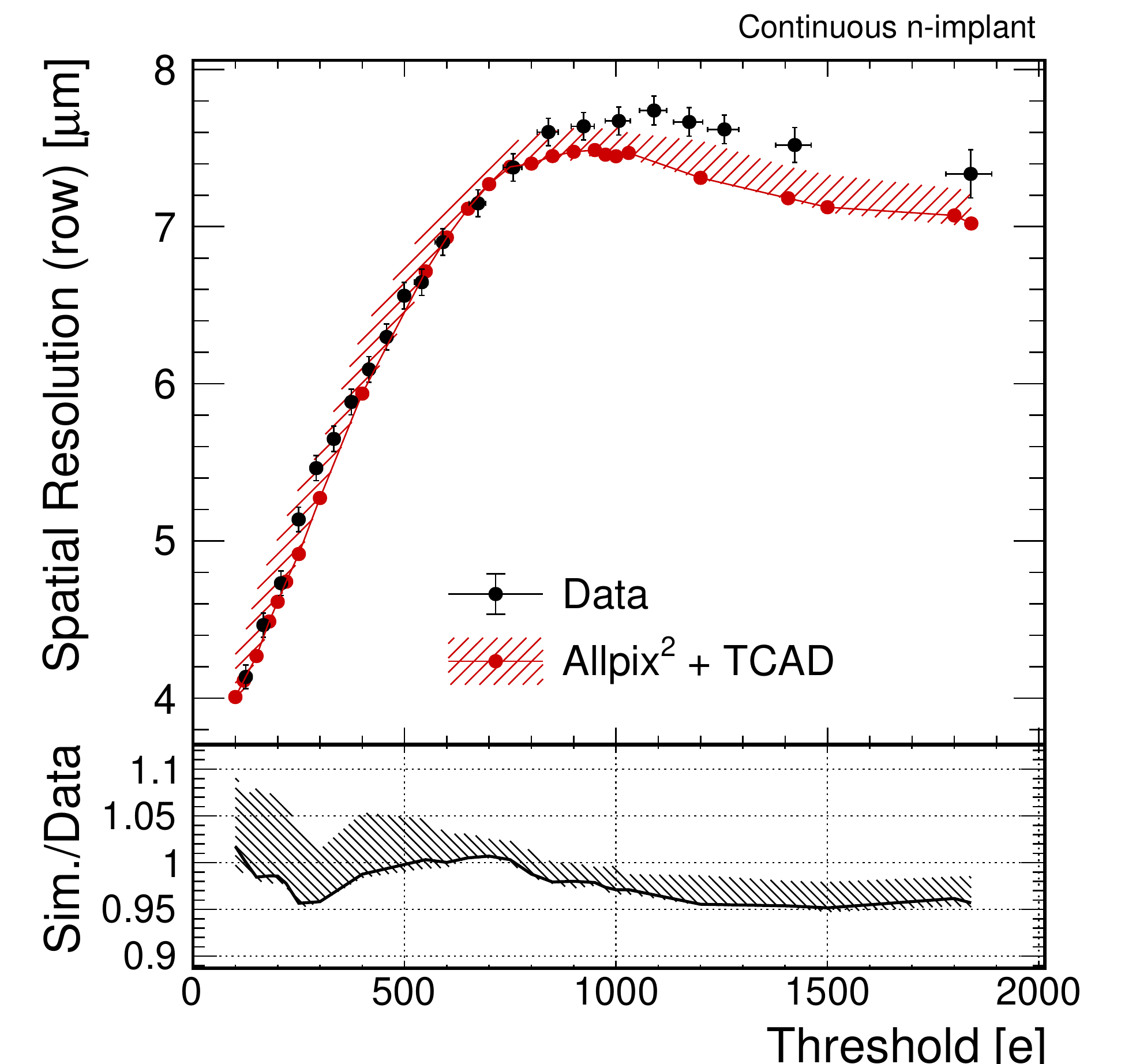}%
	    \caption{Spatial resolution as a function of the detection threshold for data and simulation.}
	   \label{fig:resolutionYSimDataThreshold}
\end{figure}

\paragraph{Threshold scan}
The spatial resolution in the row direction deteriorates with larger detection thresholds, as illustrated in Fig.~\ref{fig:resolutionYSimDataThreshold} for data and simulation.
The degradation of spatial resolution originates from the smaller cluster size at high thresholds.
An improvement in spatial resolution is observable at high detection thresholds, which originates from inefficient regions at the pixel borders, that are responsible for a smaller \textit{effective} pixel pitch from the efficient region indicated in Fig.~\ref{fig:inPxlEfficiencyThrd1000}~\cite{clictdTestbeam}.

%% file: summary.tex
A combination of electrostatic finite-element and transient Monte Carlo simulations with \apsq has been presented, and good agreement with transient 3D TCAD as well as test-beam data has been found over a wide parameter range.
The limiting factors in the simulation precision are found to pertain to the available information on the doping profiles and the front-end description of the investigated silicon device.

The simulations will be used for further development of monolithic CMOS silicon sensor with similar sensor designs, in particular in the stages of sensor optimisation and experimental assessment.
In future versions of \apsq, the modelling of the sensor front-end will be extended in order to reproduce effects arising from the electronics.

%% file: acknowledgements.tex
This work has been sponsored by the Wolfgang Gentner Programme of the German Federal Ministry of Education and Research (grant no. 05E15CHA).
The measurements leading to these results have been performed at the Test Beam Facility at DESY Hamburg (Germany), a member of the Helmholtz Association (HGF).
This project has received funding from the European Union’s Horizon 2020 research and innovation programme under grant agreement No 654168 (AIDA 2020) and No 101004761 (AIDAinnova).
This work was carried out in the framework of the CLICdp Collaboration.

%% file: credit_statement.tex
\textbf{R.~Ballabriga} Resources
\textbf{J.~Braach} Investigation
\textbf{E.~Buschmann} Investigation
\textbf{M.~Campbell} Methodology
\textbf{D.~Dannheim} Investigation, Methodology, Supervision, Writing - Review \& Editing
\textbf{K.~Dort} Formal analysis, Investigation, Visualization, Writing - Original Draft
\textbf{L. Huth} Investigation
\textbf{I.~Kremastiotis} Investigation, Resources
\textbf{J.~Kr\"oger} Investigation
\textbf{L.~Linssen} Project administration, Funding acquisition
\textbf{M.~Munker} Investigation, Methodology, Supervision, Writing - Review \& Editing
\textbf{P.~Sch\"utze} Software
\textbf{W.~Snoeys} Conceptualization, Resources
\textbf{S.~Spannagel} Investigation, Methodology, Supervision, Software, Writing - Original Draft
\textbf{T.~Vanat} Resources

%% file: paper.bbl
\begin{thebibliography}{10}
\expandafter\ifx\csname url\endcsname\relax
  \def\url#1{\texttt{#1}}\fi
\expandafter\ifx\csname urlprefix\endcsname\relax\def\urlprefix{URL }\fi
\expandafter\ifx\csname href\endcsname\relax
  \def\href#1#2{#2} \def\path#1{#1}\fi

\bibitem{apsq}
S.~Spannagel, et~al., Allpix$^2$: A modular simulation framework for silicon
  detectors, Nucl. Instr. Meth. A 901 (2018) 164 -- 172.
\newblock \href {http://dx.doi.org/10.1016/j.nima.2018.06.020}
  {\path{doi:10.1016/j.nima.2018.06.020}}.

\bibitem{allpix-hrcmos}
D.~Dannheim, et~al., {Combining TCAD and Monte Carlo Methods to Simulate CMOS
  Pixel Sensors with a Small Collection Electrode using the Allpix$^2$
  Framework}, Nucl. Instr. Meth. A 964 (2020) 163784.
\newblock \href {http://dx.doi.org/10.1016/j.nima.2020.163784}
  {\path{doi:10.1016/j.nima.2020.163784}}.

\bibitem{shockley}
W.~Shockley, Currents to conductors induced by a moving point charge, J. Appl.
  Phys. 9~(10) (1938) 635--636.
\newblock \href {http://dx.doi.org/10.1063/1.1710367}
  {\path{doi:10.1063/1.1710367}}.

\bibitem{ramo}
S.~Ramo, Currents induced by electron motion, Proc. IRE 27~(9) (1939) 584--585.
\newblock \href {http://dx.doi.org/10.1109/JRPROC.1939.228757}
  {\path{doi:10.1109/JRPROC.1939.228757}}.

\bibitem{tj-modified}
W.~Snoeys, et~al., A process modification for {CMOS} monolithic active pixel
  sensors for enhanced depletion, timing performance and radiation tolerance,
  Nucl. Instr. Meth. A 871 (2017) 90 -- 96.
\newblock \href {http://dx.doi.org/10.1016/j.nima.2017.07.046}
  {\path{doi:10.1016/j.nima.2017.07.046}}.

\bibitem{Munker:2019vdo}
M.~Munker, et~al., {Simulations of CMOS pixel sensors with a small collection
  electrode, improved for a faster charge collection and increased radiation
  tolerance}, JINST 14~(05) (2019) C05013.
\newblock \href {http://dx.doi.org/10.1088/1748-0221/14/05/C05013}
  {\path{doi:10.1088/1748-0221/14/05/C05013}}.

\bibitem{Dort:2773808}
K.~Dort, \href{https://cds.cern.ch/record/2773808}{{Test-beam and simulation
  studies for the CLICTD technology demonstrator - a monolithic CMOS pixel
  sensor with a small collection diode}}, Tech. rep., CERN, Geneva (Jun 2021).
\newline\urlprefix\url{https://cds.cern.ch/record/2773808}

\bibitem{clictd_design_characterization}
I.~Kremastiotis, et~al., {Design and Characterization of the CLICTD Pixelated
  Monolithic Sensor Chip}, IEEE Trans. Nucl. Sci. 67~(10) (2020) 2263--2272.
\newblock \href {http://dx.doi.org/10.1109/TNS.2020.3019887}
  {\path{doi:10.1109/TNS.2020.3019887}}.

\bibitem{clictdTestbeam}
R.~Ballabriga, et~al., {Test-beam characterisation of the CLICTD technology
  demonstrator - A small collection electrode high-resistivity CMOS pixel
  sensor with simultaneous time and energy measurement}, Nucl. Instr. Meth. A
  1006 (2021) 165396.
\newblock \href {http://dx.doi.org/https://doi.org/10.1016/j.nima.2021.165396}
  {\path{doi:https://doi.org/10.1016/j.nima.2021.165396}}.

\bibitem{sentaurus-website}
\href{https://www.synopsys.com}{Synopsys sentaurus device}, accessed 06~2021.
\newline\urlprefix\url{https://www.synopsys.com}

\bibitem{Meroli:2011zzb}
S.~Meroli, D.~Passeri, L.~Servoli, {Energy loss measurement for charged
  particles in very thin silicon layers}, JINST 6 (2011) P06013.
\newblock \href {http://dx.doi.org/10.1088/1748-0221/6/06/P06013}
  {\path{doi:10.1088/1748-0221/6/06/P06013}}.

\bibitem{geant4}
S.~Agostinelli, et~al., Geant4 -- a simulation toolkit, Nucl. Instr. Meth. A
  506~(3) (2003) 250 -- 303.
\newblock \href {http://dx.doi.org/10.1016/S0168-9002(03)01368-8}
  {\path{doi:10.1016/S0168-9002(03)01368-8}}.

\bibitem{geant4-2}
J.~Allison, et~al., Geant4 developments and applications, IEEE Trans. Nucl.
  Sci. 53~(1) (2006) 270--278.
\newblock \href {http://dx.doi.org/10.1109/TNS.2006.869826}
  {\path{doi:10.1109/TNS.2006.869826}}.

\bibitem{planecondenser}
W.~Riegler, G.~A. Rinella, Point charge potential and weighting field of a
  pixel or pad in a plane condenser, Nucl. Instr. Meth. A 767 (2014) 267 --
  270.
\newblock \href {http://dx.doi.org/10.1016/j.nima.2014.08.044}
  {\path{doi:10.1016/j.nima.2014.08.044}}.

\bibitem{fossum-lee}
J.~Fossum, D.~Lee, A physical model for the dependence of carrier lifetime on
  doping density in nondegenerate silicon, Solid-State Electronics 25~(8)
  (1982) 741 -- 747.
\newblock \href
  {http://dx.doi.org/https://doi.org/10.1016/0038-1101(82)90203-9}
  {\path{doi:https://doi.org/10.1016/0038-1101(82)90203-9}}.

\bibitem{fossum}
J.~G. Fossum, Computer-aided numerical analysis of silicon solar cells,
  Solid-State Electronics 19~(4) (1976) 269 -- 277.
\newblock \href
  {http://dx.doi.org/https://doi.org/10.1016/0038-1101(76)90022-8}
  {\path{doi:https://doi.org/10.1016/0038-1101(76)90022-8}}.

\bibitem{FOSSUM1983569}
J.~Fossum, R.~Mertens, D.~Lee, J.~Nijs, Carrier recombination and lifetime in
  highly doped silicon, Solid-State Electronics 26~(6) (1983) 569--576.
\newblock \href
  {http://dx.doi.org/https://doi.org/10.1016/0038-1101(83)90173-9}
  {\path{doi:https://doi.org/10.1016/0038-1101(83)90173-9}}.

\bibitem{augerKerr2002}
M.~Kerr, A.~Cuevas, {General parametrization of Auger Recombination in
  crystalline silicon}, J. Appl. Phys. 91.
\newblock \href {http://dx.doi.org/10.1063/1.1432476}
  {\path{doi:10.1063/1.1432476}}.

\bibitem{DziewiorSchmid}
J.~{Dziewior}, W.~{Schmid}, {Auger coefficients for highly doped and highly
  excited silicon}, Appl. Phys. Lett. 31~(5) (1977) 346.
\newblock \href {http://dx.doi.org/10.1063/1.89694}
  {\path{doi:10.1063/1.89694}}.

\bibitem{masetti}
G.~Masetti, M.~Severi, S.~Solmi, Modeling of carrier mobility against carrier
  concentration in arsenic-, phosphorus-, and boron-doped silicon, IEEE Trans.
  Elec. Dev. 30~(7) (1983) 764--769.
\newblock \href {http://dx.doi.org/10.1109/T-ED.1983.21207}
  {\path{doi:10.1109/T-ED.1983.21207}}.

\bibitem{canali}
C.~Canali, G.~Majni, R.~Minder, G.~Ottaviani, Electron and hole drift velocity
  measurements in silicon and their empirical relation to electric field and
  temperature, IEEE Trans. Elec. Dev. 22~(11) (1975) 1045--1047.
\newblock \href {http://dx.doi.org/10.1109/T-ED.1975.18267}
  {\path{doi:10.1109/T-ED.1975.18267}}.

\bibitem{apsq_manual}
\href{https://cern.ch/allpix-squared/usermanual/allpix-manual.pdf}{\apsq user
  manual, version 2.0.2}, accessed 10~2021.
\newline\urlprefix\url{https://cern.ch/allpix-squared/usermanual/allpix-manual.pdf}

\bibitem{fehlberg}
E.~Fehlberg, \href{https://ntrs.nasa.gov/search.jsp?R=19690021375}{Low-order
  classical {Runge-Kutta} formulas with stepsize control and their application
  to some heat transfer problems}, NASA Technical Report NASA-TR-R-315, NASA
  (1969).
\newline\urlprefix\url{https://ntrs.nasa.gov/search.jsp?R=19690021375}

\bibitem{Diener:2018qap}
R.~Diener, et~al., {The DESY II Test Beam Facility}, {Nucl. Instr. Meth. A} 922
  (2019) 265--286.
\newblock \href {http://dx.doi.org/10.1016/j.nima.2018.11.133}
  {\path{doi:10.1016/j.nima.2018.11.133}}.

\bibitem{Jansen:2016bkd}
H.~Jansen, et~al., {Performance of the EUDET-type beam telescopes}, EPJ Tech.
  Instrum. 3~(1) (2016) 7.
\newblock \href {http://dx.doi.org/10.1140/epjti/s40485-016-0033-2}
  {\path{doi:10.1140/epjti/s40485-016-0033-2}}.

\bibitem{Poikela_2014}
T.~Poikela, et~al., Timepix3: a 65k channel hybrid pixel readout chip with
  simultaneous {ToA}/{ToT} and sparse readout, JINST 9~(05) (2014)
  C05013--C05013.
\newblock \href {http://dx.doi.org/10.1088/1748-0221/9/05/c05013}
  {\path{doi:10.1088/1748-0221/9/05/c05013}}.

\bibitem{corry_paper}
D.~Dannheim, et~al., {Corryvreckan: A Modular 4D Track Reconstruction and
  Analysis Software for Test Beam Data}, JINST 16~(03) (2021) P03008.
\newblock \href {http://dx.doi.org/10.1088/1748-0221/16/03/p03008}
  {\path{doi:10.1088/1748-0221/16/03/p03008}}.

\bibitem{Akiba:2011vn}
K.~Akiba, et~al., {Charged Particle Tracking with the Timepix ASIC}, Nucl.
  Instr. Meth. A 661 (2012) 31--49.
\newblock \href {http://dx.doi.org/10.1016/j.nima.2011.09.021}
  {\path{doi:10.1016/j.nima.2011.09.021}}.

\bibitem{Blobel:2006yi}
V.~Blobel, {A new fast track-fit algorithm based on broken lines}, Nucl. Instr.
  Meth. A 566 (2006) 14--17.
\newblock \href {http://dx.doi.org/10.1016/j.nima.2006.05.156}
  {\path{doi:10.1016/j.nima.2006.05.156}}.

\bibitem{resolutionSimulator}
S.~Spannagel, H.~Jansen,
  \href{https://github.com/simonspa/resolution-simulator}{Gbl track resolution
  calculator v2.0, 2016}.
\newblock \href {http://dx.doi.org/doi:10.5281/zenodo.48795}
  {\path{doi:doi:10.5281/zenodo.48795}}.
\newline\urlprefix\url{https://github.com/simonspa/resolution-simulator}

\bibitem{thesis-jacobus}
J.~W. van Hoorne, \href{https://cds.cern.ch/record/2119197}{{Study and
  Development of a novel Silicon Pixel Detector for the Upgrade of the ALICE
  Inner Tracking System}}, {PhD thesis}, Technische Universit\"at Wien (Nov
  2015).
\newline\urlprefix\url{https://cds.cern.ch/record/2119197}

\end{thebibliography}
